\documentclass[]{article}
\usepackage{graphicx}
\usepackage{amsmath}
\usepackage{caption}
\usepackage{float}
\floatstyle{plaintop}
\restylefloat{table}

\begin{document}

\title{Analysis of terahertz generation by beamlet superposition}
\author{Koustuban Ravi,$^{1,2,*}$ B. K. Ofori-Okai,$^{3}$ Keith A. Nelson,$^{3}$ \\
and Franz X. K\"artner $^{1,2,4}$}

\maketitle

\noindent 1. Center for free-electron laser science, DESY, Notkestra$\beta$e 85, Hamburg 22607, Germany\\
2. Research Laboratory of Electronics, Massachusetts Institute of Technology, Cambridge MA 02139, USA\\
3. Department of Chemistry, Massachusetts Institute of Technology, Cambridge MA 02139, USA\\
4. Department of Physics, University of Hamburg, Hamburg 22761, Germany
Email : $^*$koustuban@alum.mit.edu 

\begin{abstract}
 A theory of terahertz generation using a superposition of beamlets is developed. It is shown how such an arrangement produces a distortion-free tilted pulse front. We analytically show how a superposition of beamlets produces terahertz radiation with greater efficiency and spatial homogeneity compared to tilted pulse fronts generated by diffraction gratings. The advantages are particularly notable for large pump bandwidths and beam sizes, suggesting better performance in the presence of cascading effects and for high energy pumping. Closed-form expressions for terahertz spectra and transients in three spatial dimensions are derived. Conditions for obtaining performance better than conventional tilted pulse fronts and bounds for optimal pump parameters are furnished.
\end{abstract}

\section{Introduction}
Terahertz pulses with large electric field strengths ($\sim0.1-10$ MV cm$^{-1}$) are attractive for a number of fundamental scientific investigations \cite{kampfrath2013,schubert2014} and technological applications. Compact particle acceleration \cite{nanni2015,palfalvi2014,arya2016,ronnyhuang16,zhang2018,lemery2018}, electron microscopy, and electron-beam diagnostics \cite{kealhofer2016,ropers2016} have been predicted to benefit from advances in terahertz generation. While many modalities for terahertz generation exist \cite{gallerano2004,gold1997}, optical rectification of sub-picosecond near-infrared (NIR) frequency pulses has achieved great success. Using lithium niobate as the nonlinear medium, terahertz pulses have been produced with percent-level energy conversion efficiencies \cite{huang2015} and millijoule-level pulse energies \cite{fulop2014}. The highest conversion efficiencies are possible by cascaded difference frequency generation where a single pump photon is repeatedly down converted in energy to yield multiple terahertz photons\cite{cronin2004}. 
 
Compatibility with off-the-shelf titanium sapphire and ytterbium laser technology has made terahertz generation using tilted pulse fronts \cite{hebling02,yeh2007,fulop10,fulop2011,hirori} in lithium niobate ubiquitous. Tilted pulse fronts are required for phase matching because of the large difference in the NIR and terahertz refractive indices of lithium niobate ($n_{\textrm{NIR}} \sim 2.2$ and $n_{\textrm{THz}} \sim 5$). In practice, the pulse-front tilt is produced using a diffraction grating \cite{Hebling1996}. As each frequency component of the pump spectrum, $\omega_{i}$, propagates away from the grating at a different angle, $\theta_{i}$, a tilted pulse front results. The generated tilted pulse front is then imaged into a lithium niobate prism. 

However, the inherently large group-velocity dispersion due to angular dispersion (GVD-AD) limits terahertz generation for large pump bandwidths. Due to GVD-AD, the angular frequency separation, $d\theta/d\omega$, is not constant. This prevents phase matching between pump and terahertz frequency components across the entire pump pulse bandwidth. Furthermore, imaging tilted pulse fronts produced with diffraction gratings is a challenging task when the pump pulse has a large bandwidth. Even if the initial pump bandwidth is relatively narrow, bandwidth limiting issues arise when significant cascading broadens the pump spectrum \cite{ravi14,ravi15}. 

One proposed approach to circumvent these limitations is to use a tilted pulse front assembled by a superposition of small, discrete, time-delayed pump beamlets. This was demonstrated in \cite{ofori2016thz}, and variants of the approach were studied in \cite{avetisyan2017,avetisyan2017_2,palfalvi2017,toth2019,nugraha2019}. Conceptually, the small beamlet size in one of the transverse directions (e.g. $x$) results in a large, uniform spread in transverse momentum ($k_x$). The transverse momentum spread enables non-collinear phase matching of the pump and terahertz radiation. Each beamlet generates terahertz radiation at a Cherenkov angle, $\gamma =\cos^{-1}(n_{NIR}/n_{THz})$, and the superposition of the generated radiation from various beamlets forms a terahertz plane wave.

Most importantly, as the angular spread is uniform for all pump spectral components, phase matching may be maintained across the entire pump pulse bandwidth. This allows the approach to potentially circumvent limitations due to bandwidth and cascading effects \cite{ravi2016}. However, a detailed theoretical investigation has not yet been performed to explore this approach comprehensively.

Here we present an analytic formulation in three spatial dimensions in the undepleted limit that establishes the advantages offered by beamlet superposition compared to tilted pulse fronts generated by diffraction gratings. We consider the effects of absorption, diffraction of optical and terahertz beams, beam curvature, and interaction within and between pump beamlets. However, we do not consider the effects of pump depletion and associated cascading effects. While neglecting cascading effects could change quantitative predictions significantly, important qualitative understanding can still be obtained in the undepleted limit.

We derive closed-form expressions which suggest terahertz spectra with reduced spatial inhomogeneities, higher frequencies, and higher conversion efficiencies compared to terahertz radiation generated by diffraction-grating-based tilted pulse fronts (DG-TPF). Conditions for obtaining conversion efficiencies better than DG-TPFs and the limitations of the approach are discussed. Conversion efficiencies obtained from beamlet superposition are found to be superior for larger pump bandwidths and beam sizes. These findings also suggest greater insensitivity of beamlet superposition to cascading effects. It is worthwhile to point out the difference between the terms ``beamlet'' and ``beam''. Here, beam size corresponds to the spatial extent of the superposition of all beamlets.

We first present the theoretical formulation and physical mechanisms in Section 2. Next, we show results comparing terahertz generation by beamlet superposition and DG-TPFs in Section 3. Finally, conclusions are given in Section 4. Appendices are provided for the sake of completeness.

\section{Theory}
\subsection{Tilted pulse fronts by beamlet superposition}
\begin{figure}
\begin{center}
\scalebox{0.3}[0.3]{\includegraphics{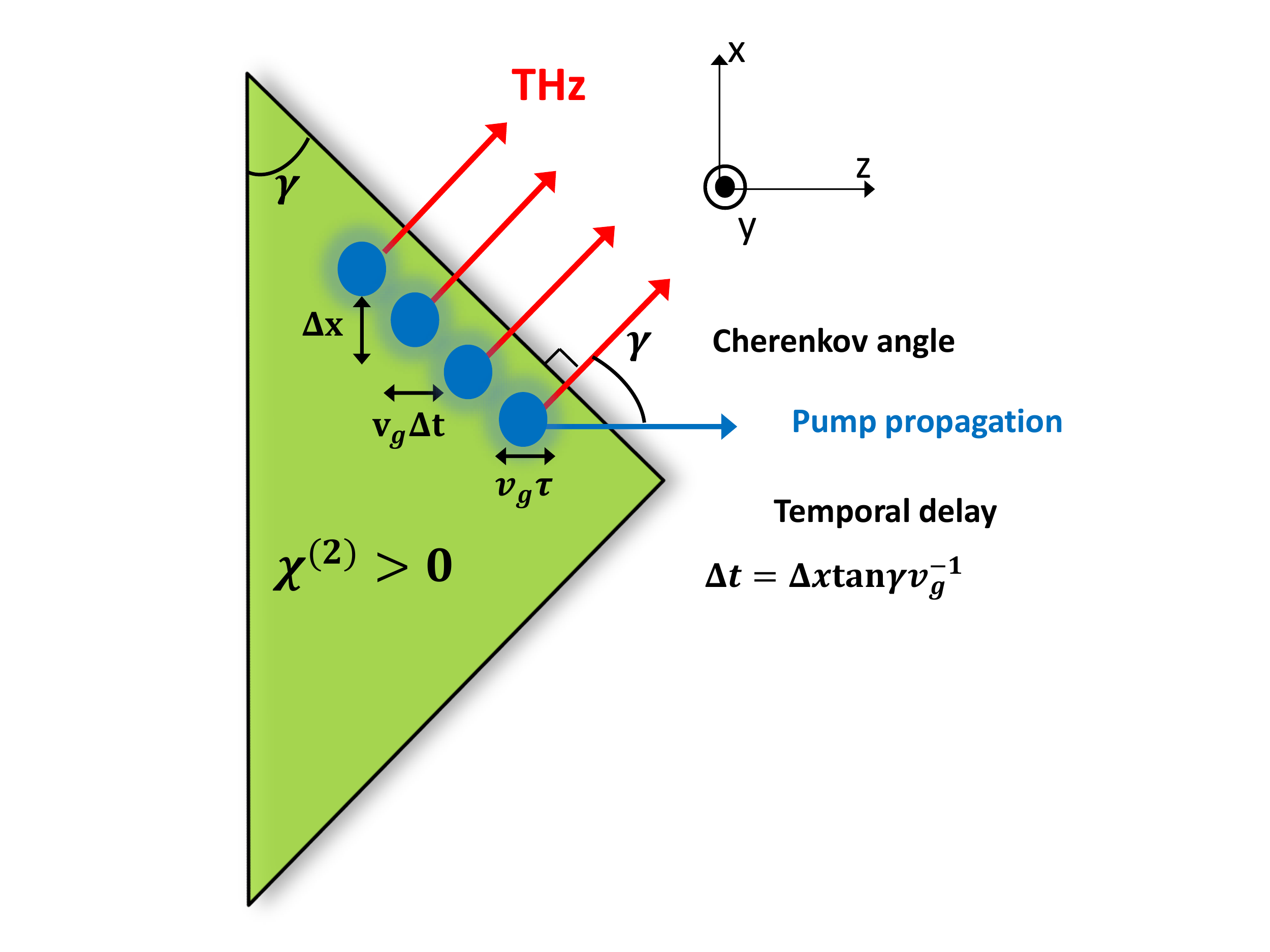}}
\caption{\label{fig1s}Schematic of terahertz generation using a superposition of beamlets of $1/e^{2}$ duration $\tau$, separated transversely by a distance $\Delta x$ and longitudinally by $v_g\Delta t$. Each beamlet emits radiation at the Cherenkov angle, $\gamma$, which sets the condition for coherent superposition of radiation to be $v_g |\Delta t| = |\Delta x|\tan\gamma$.}
\end{center}
\end{figure}

\begin{figure}
\begin{center}
\includegraphics[scale=0.35]{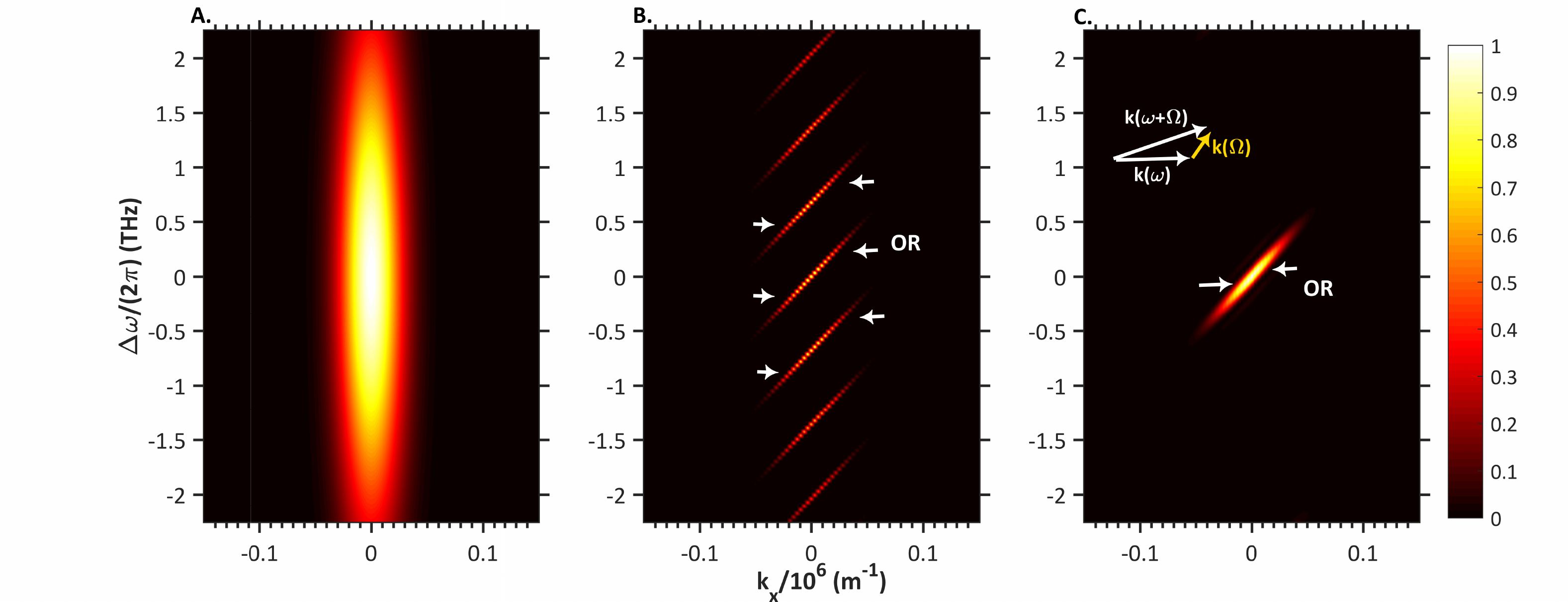}
\caption{\label{fig1}(a) $|E(k_x,\omega)|^2$ of a single beamlet with $\tau=50$ fs and $\sigma=50$ $\mu$m shows a Gaussian spread in transverse momentum and frequency. (b) $N=20$ beamlets separated by $100$ $\mu$m appear as a series of oblique lines in $(\omega,k_x)$- space. The slope of the lines are $v_g\text{tan}\gamma$, where $\gamma$ is the pulse-front tilt angle. The spacing between lines is inversely proportional to the spacing between successive beamlets $\Delta x$ in real space. The white arrows are used to delineate the fact that optical rectification occurs within the $\omega-k_x$ distribution specified by each of the oblique lines. (c) When the spacing between lines is reduced to $\Delta x =\sigma/2=25$ $\mu$m,  only a single oblique line falls within the Gaussian spread of $k_x$.}
\end{center}
\end{figure}

Figure \ref{fig1} depicts the system. A series of pump beamlets, each with $1/e^{2}$ duration $\tau$, propagate with group velocity $v_g=c/n_{g}$ in the $z$-direction in an electro-optic crystal  with a group refractive index, $n_{g}$, and a second-order nonlinear coefficient, $\chi^{(2)}$. The geometry of the crystal is assumed to be similar to that employed in previous experiments \cite{huang2015}. The beamlets are separated along the $x$-direction by an amount $\Delta x$ and by an amount $\Delta z$ in the $z-$direction. From Fig. \ref{fig1s}, beamlets at larger transverse positions are located at smaller longitudinal positions, so $\Delta z=-\Delta x\tan\gamma$.  Furthermore, they are temporally separated in increments of $|\Delta t|= |\Delta x| \text{tan}\gamma/v_g$.

In the frequency-domain, the total electric field of all beamlets can be represented as,

\begin{multline}
E(x,y,z,\omega)=\sqrt{\frac{\sigma}{\sigma(z)}}\sum_{q}E_q(\omega)e^{-\frac{y^2}{w_y^2}}e^{-\frac{(x-x_q)^{2}}{\sigma(z)^2}}\times \\
e^{-j\Delta\omega v_g^{-1}(z-z_q)}e^{-jk_m\Delta\omega^2z}e^{-\frac{jk(\omega)(x-x_q)^{2}}{2R(z)}}e^{-j\frac{\phi^{(2)}\Delta\omega^2}{2}}.\label{eq:5a}
\end{multline}

\begin{align}
E_q(\omega) = E_qe^{-\frac{\Delta\omega^2\tau^2}{4}}\label{eq:5d}\\
\sigma(z) = \sigma\bigg[1+\left(\frac{z-z_0}{z_R}\right)^2\bigg]^{1/2}\label{eq:5b}\\
R^{-1}(z) = \frac{z-z_0}{ (z-z_0)^2+z_R^2}\label{eq:5c}
\end{align}

The $y-$dependent Gaussian term in Eq. (\ref{eq:5a}) delineates the spatial profile of the pump field in the $y$-direction (out-of-plane direction). The beam size, $w_y$, is similar to those in DG-TPFs. Furthermore, each beamlet has an $1/e^{2}$ radius defined by $\sigma$ in the $x$-direction, where $\sigma$ ($\leq100$ $\mu$m) is significantly smaller than the typical beam size ($\geq$ mm) used for terahertz generation by DG-TPFs.
 
In Eq. (\ref{eq:5a}), $x_q$ is the position of the $q^{th}$ beamlet, and $(z-z_q)/v_g$ is the corresponding temporal delay. 

Material dispersion is accounted for by the $k_m\Delta\omega^2$ term in Eq. (\ref{eq:5a}), where $\Delta\omega=\omega-\omega_0$ represents the displacement from the central angular frequency $\omega_0$ of the pump.

The last two terms of Eq. (\ref{eq:5a}) represent phases due to a finite radius of curvature, $R(z)$ and an externally imparted group-delay dispersion, $\phi^{(2)}$, respectively.

Equation (\ref{eq:5d}) represents the Gaussian spectrum centered about angular frequency $\omega_0$. The small size of beamlets in the $x$-direction results in a Rayleigh length on the order of 1 cm. This is comparable to the typical terahertz absorption length in lithium niobate at cryogenic temperatures. As such, we also consider longitudinal variations of $\sigma$ and radius of curvature $R(z)$ in Eqs. (\ref{eq:5b})-(\ref{eq:5c}). Here, $z_0$ represents the location of the beam waist, while $z_R=k(\omega_0)n(\omega_0)\sigma^2/2$ corresponds to the Rayleigh length. The $(\sigma/\sigma(z))^{1/2}$ prefactor in Eq. (\ref{eq:5a}) represents the conservation of energy as the size of beamlets change along $z$.

It is useful to visualize the distribution of the field in $(k_x,\omega)$-space, where $k_x$ is the $x$-directed transverse momentum. Spatial Fourier transformation of Eq. (\ref{eq:5a}) gives

\begin{align}
E(k_x,z,\omega)|_{y=0} = \frac{\sigma}{2\sqrt{\pi}}A(\omega)e^{-k_{x}^{2}\sigma^{2}/4}\sum_{q}\text{e}^{jq(k_{x}\Delta x+\Delta\omega \Delta z/v_{g})}, \label{eq:6a}\\
=\frac{\sigma}{2\sqrt{\pi}}A(\omega)\text{e}^{-k_{x}^{2}\sigma^{2}/4}\frac{\text{sin}[\frac{N}{2}\Delta x(k_{x}+\Delta\omega \frac{\Delta z}{\Delta x}/v_{g})]}{\text{sin}[\frac{1}{2}\Delta x(k_{x} +\Delta\omega \frac{\Delta z}{\Delta x}/v_{g}]}.\label{eq:6b}
\end{align}

\noindent In Eq. (\ref{eq:6a}), $E_q(\omega)=A(\omega)$ without loss of generality. For illustrative purposes, we ignore the radius of curvature and variations along the $y$-direction, since the key physics only involves the $(x,z)$-plane.

Figure \ref{fig1} shows normalized (within each panel) values of $|E(k_x,\omega,z)|^2$ for various cases. In Fig. \ref{fig1}(a), there is a single beamlet of $1/e^2$ pulse duration $\tau=50$ fs and $1/e^{2}$ beamlet radius $\sigma=50\,\mu$m. The small size of the beamlet produces a large transverse momentum spread while the large bandwidth produces a significant spread in frequency. This represents the behavior of the Gaussian prefactor in Eqs. (\ref{eq:6a})-(\ref{eq:6b}). It is worth noting that the full-width at half-maximum spectral bandwidth of a 50$\,$fs pulse is about 7.5$\,$THz. However, only $5\,$THz is shown in Fig. \ref{fig1}(a) for illustrative reasons.

Figure \ref{fig1}(b) shows the case of $N=20$ beamlets separated by a distance $\Delta x=2\sigma=100$ $\mu$m. Here, $E(\omega,k_x,z)$ appears as a set of parallel, oblique lines. These lines represent points where $k_x=\textit{l}\pi/\Delta x-\Delta\omega\tan\gamma/v_g$ for some integer $l$. This condition maximizes the ratio of the sinusoidal functions in Eq. (\ref{eq:6b}). Note that the spacing between lines is inversely proportional to the transverse separation of beamlets, $\Delta x$. 

In Fig. \ref{fig1}(c), the spacing between beamlets is reduced to $\Delta x =\sigma/2=25$ $\mu$m. Here, only one line is visible in the region of the $\Delta\omega-k_x$ plane depicted in Fig. \ref{fig1}. Higher-order oblique lines are still present due to the ratio of sinusoidal functions in Eq. (\ref{eq:6b}). However, they are of much lower intensity compared to the oblique line passing through $(k_x,\Delta\omega)=(0,0)$. This illustrates that when the spacing between beamlets is sufficiently small, only one oblique line falls within $1/e^{2}$ of the Gaussian prefactor in Eqs. (\ref{eq:6a})-(\ref{eq:6b}). Note that the apparent reduction in the spectral bandwidth of the ensemble of beamlets in Fig. \ref{fig1}(c) is a consequence of the Fourier transform between spatial and transverse momentum domains. Physically speaking, each individual beamlet still possesses the same spectral bandwidth. 

The $k_x$-separation between consecutive oblique lines is $\pi/\Delta x$ for $\Delta\omega=0$. This suggests that an optimal beamlet spacing can be found by setting $(\sigma k_x)^2/4= 2$ in the Gaussian pre-factor of Eq. (\ref{eq:6a}). For the $k_x$-separation of $\pi/\Delta x$, this yields the following bound on $\Delta x$ in Eq. (\ref{dx_c}):

\begin{gather}
\Delta x \leq \frac{\pi\sigma}{2\sqrt{2}}\approx \sigma. \label{dx_c}
\end{gather}

In establishing the above criterion, we have assumed Gaussian spatial profiles for the beamlets. However, in experimental situations such as in the case of an echelon \cite{ofori2016thz}, the beamlets are better represented by top-hat spatial profiles. In such a case, the requirement for proximity could be alleviated. 

\subsection{Comparison to tilted pulse fronts from diffraction gratings}

Equations (\ref{TPF_ex}) and (\ref{TPF_kx}) represent the fields due to a diffraction grating in the $(x,\omega)$- and $(k_x,\omega)$-spaces respectively,

\begin{gather}
E(x,z,\omega)_{DG} =E_0e^{-\frac{\Delta\omega^2\tau^2}{4}}e^{-x^2/w^2}e^{-j\phi_{x}x}e^{-j\phi_{z}z},\label{TPF_ex}\\
E(k_x,z,\omega)_{DG} = \frac{w}{2\sqrt{\pi}}E_0e^{-\frac{\Delta\omega^2\tau^2}{4}}e^{-\left(k_x-\phi_{x}\right)^2w^2/4}e^{-j\phi_{z}z},\label{TPF_kx}
\end{gather}

\noindent where

\begin{align}
\phi_{x} = k(\omega_0)\beta\Delta\omega+k_T\Delta\omega^2,\label{phi_x}\\
\phi_{z} = \Delta\omega /v_g.
\end{align}

In Eq. (\ref{phi_x}), $\beta=d\theta/d\omega$ is the angular dispersion term and $k_T$ denotes GVD-AD. Equation (\ref{TPF_kx}) indicates that the distribution of the field in $(k_x,\omega)$-space is maximized at $k_x = \phi_{x}$.

\begin{figure}
\begin{center}
{\includegraphics[scale=0.3]{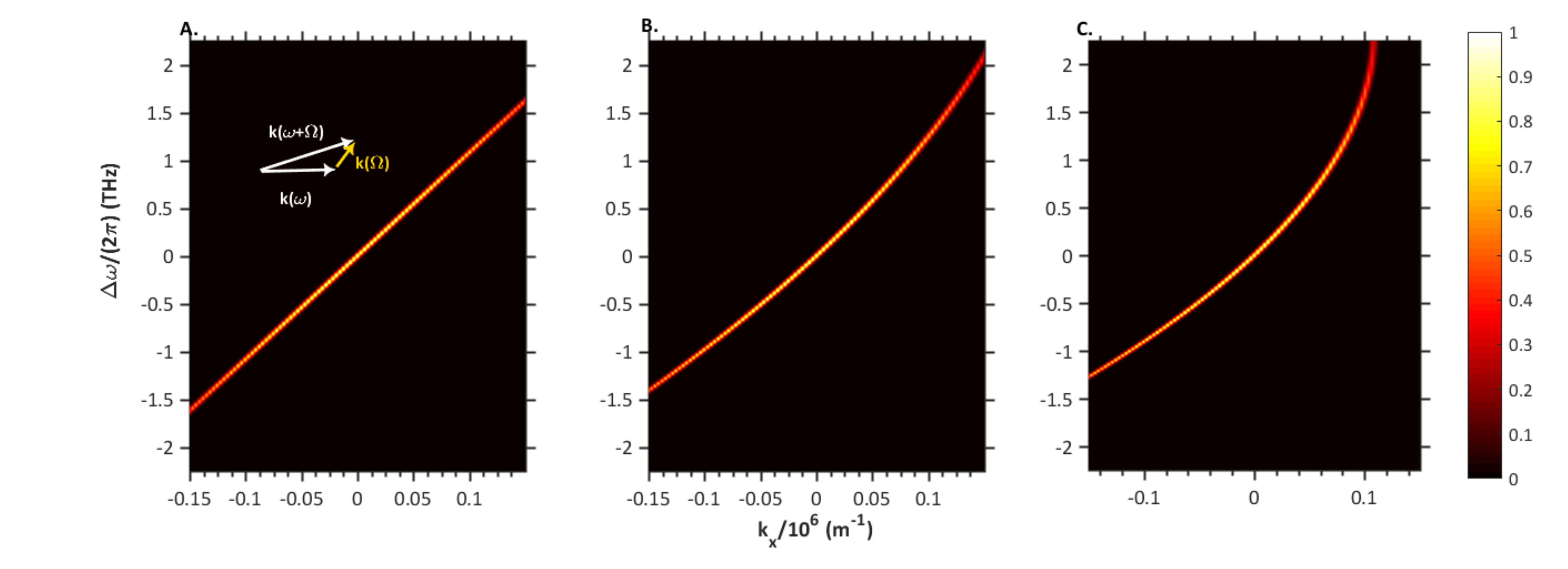}}
\caption{\label{fig2} Plots of the field from Eq. (\ref{TPF_kx}) in the ($k_x,\Delta\omega$)-plane for a $\tau$=50 \,fs, $w=0.5$\,mm pulse for different values of group-velocity dispersion due to angular dispersion. (a) $k_T=0$ (b) $k_T=-25 \times10^{-23}\,\text{s}^2\text{m}^{-1}$ (c) $-50\times 10^{-23}\,\text{s}^2$ .The first case represents a distortion-free tilted pulse front. The subsequent cases do not satisfy phase matching conditions over the entire bandwidth of the incident pulse due to a finite value of group-velocity dispersion due to angular dispersion, $k_T$.}
\end{center}
\end{figure}

Figure \ref{fig2}(a) shows the spectral distribution for a tilted pulse front in the absence of GVD-AD ($k_T=0$). For a tilt angle of $\gamma$, it can be shown by examining the line of constant phase from from Eq. (\ref{TPF_ex}) that $k(\omega_0)\beta v_g=\text{tan}\gamma$. In comparison with the previous discussion on beamlet superposition, it is clear that an appropriately designed series of delayed beamlets (Fig. \ref{fig1}(c)) behaves like a distortion-free tilted pulse front in the ideal limit. Additionally, for a grating based TPF, the distributions have a narrower spread due to the large value of beam size $w$ in contrast to the beamlet size $\sigma$. This is one advantage of tilted pulse fronts formed by gratings: all the energy is concentrated at the desired transverse momentum value in the ideal limit. However, when $k_T\neq0$ (Figs. \ref{fig2}(b)-\ref{fig2}(c)), one obtains a nonlinear distribution in $(k_x, \Delta \omega)$-space for large $\Delta\omega$, illustrating that for equal increments in frequency, the angular separation is unequal. A consequence of this nonlinear distribution is that phase matching conditions are not satisfied across the bandwidth of the pulse. In Figs. \ref{fig2}(b)-\ref{fig2}(c), the values of $k_T$ are purposely exaggerated for illustrative purposes. However, we use experimentally relevant values of $k_T$ in our calculations (See Table \ref{param_list}).

\subsection{Nonlinear polarization}

To obtain expressions for the generated terahertz spectra and transients, we consider the second-order nonlinear polarization term (see appendix for derivation) at the terahertz angular frequency $\Omega$ for a nonlinear material with an effective second-order susceptibility $\chi^{(2)}$. The nonlinear polarization drives the terahertz generation process. The general expression for the case of a superposition of beamlets is shown in Eq. (\ref{eq:7_1a}):

\begin{multline}
P_{THz}(x,y,z,\Omega)=
\varepsilon_0\chi^{(2)}\frac{\sigma}{\sigma(z)}\frac{\sqrt{2\pi}}{\tau}
e^{-\frac{2y^2}{w_y^2}}
e^{-\frac{\Omega^2\tau_1^2}{8}} e^{-j\frac{\Omega z}{v_g}}\times \\
\sum_{p}\sum_{q}G_{p,q}e^{j\Omega \langle z_{p,q}\rangle /v_{g}}S_{p,q}(x,z,\Omega). \label{eq:7_1a}
\end{multline}
\noindent The different components of this expression are discussed below. A list of all variables are provided in Table \ref{var_list} in the appendix for quick reference.


\subsubsection{Bandwidth effects}

\noindent The first and second Gaussian terms in Eq. (\ref{eq:7_1a}) represent the spatial profile in the $y$-direction and the terahertz spectral content respectively. The effective duration of the nonlinear polarization is $\tau_1/\sqrt{2}$ where
\begin{align}
\tau_1^2  = \tau^2 + \frac{16\beta''^2}{\tau^2}\label{eq:7_1c}, \\
\beta'' = \frac{\phi^{(2)}}{2} +k_m z.\label{eq:7_1b}
\end{align} 

\noindent In Eq. (\ref{eq:7_1b}), $\beta''$ accounts for the effects of any group-delay dispersion, $\phi^{(2)}$, and group-velocity dispersion due to material dispersion (GVD-MD) $k_{m}$. These effects increase the effective pump pulse duration  and reduce the generated terahertz field. The $e^{-j\Omega z/v_g }$ term in Eq. (\ref{eq:7_1a}) indicates that the nonlinear polarization propagates with group velocity $v_{g}$ in the $z$-direction. 

\subsubsection{Interaction between beamlets}

\noindent The summation over indices $p,q$ in Eq. (\ref{eq:7_1a}) delineates the combined contribution of nonlinear interactions between each pair of beamlets $p,q$. The strength of each of these interactions is given by $G_{p,q}$,

\begin{gather}
G_{p,q}= \sum_p\sum_q E_p E_q e^{-\Delta x_{p,q}^2/2\sigma^2}e^{-\Delta z_{p,q}^2/2(v_g\tau_1)^2}.\label{F_mk}
\end{gather}

\noindent $G_{p,q}$ is proportional to the spectral amplitudes, $E_p$ and $E_q$, of the $p^{th}$ and $q^{th}$ beamlets, respectively, and exhibits a Gaussian decay with respect to the transverse separation $\Delta x_{p,q}= x_p-x_q$ and longitudinal separation $\Delta z_{p,q}= z_p-z_q$. The corresponding $1/e^{2}$ values are the transverse beamlet size, $\sigma$, and the spatial pulse length, $v_g\tau$, respectively. This is intuitive in that the interaction between beamlets should weaken when their spatial separation $\Delta x_{p,q}$ exceeds the transverse beamlet size, or the temporal separation, $\Delta z_{p,q}/v_g$, significantly exceeds the pulse duration. The complex exponential term following $G_{p,q}$ in Eq. (\ref{F_mk}), delineates a delay proportional to the average longitudinal position $\langle z_{p,q}\rangle=(z_p+z_q)/2$. 

Finally, $S_{p,q} (x,z,\Omega)$, represents the transverse spatial variation of the nonlinear polarization in the $x$-direction,
\begin{gather}
S_{p,q}(x,z,\Omega)=e^{-\frac{2(x-\langle x_{p,q}\rangle)^2}{\sigma^2(z)}}
e^{-j\frac{\Omega(x-\langle x_{p,q}\rangle)^2}{2R(z)v_g}}.\label{source_term}
\end{gather}
$S_{p,q}$ suggests that the terahertz radiation generated by the interaction between the beamlet pair - $(p,q)$ is centered at their average transverse position given by $\langle x_{p,q} \rangle =(x_p+x_q)/2$.

\begin{figure}
\centering
\scalebox{0.43}[0.43]{\includegraphics{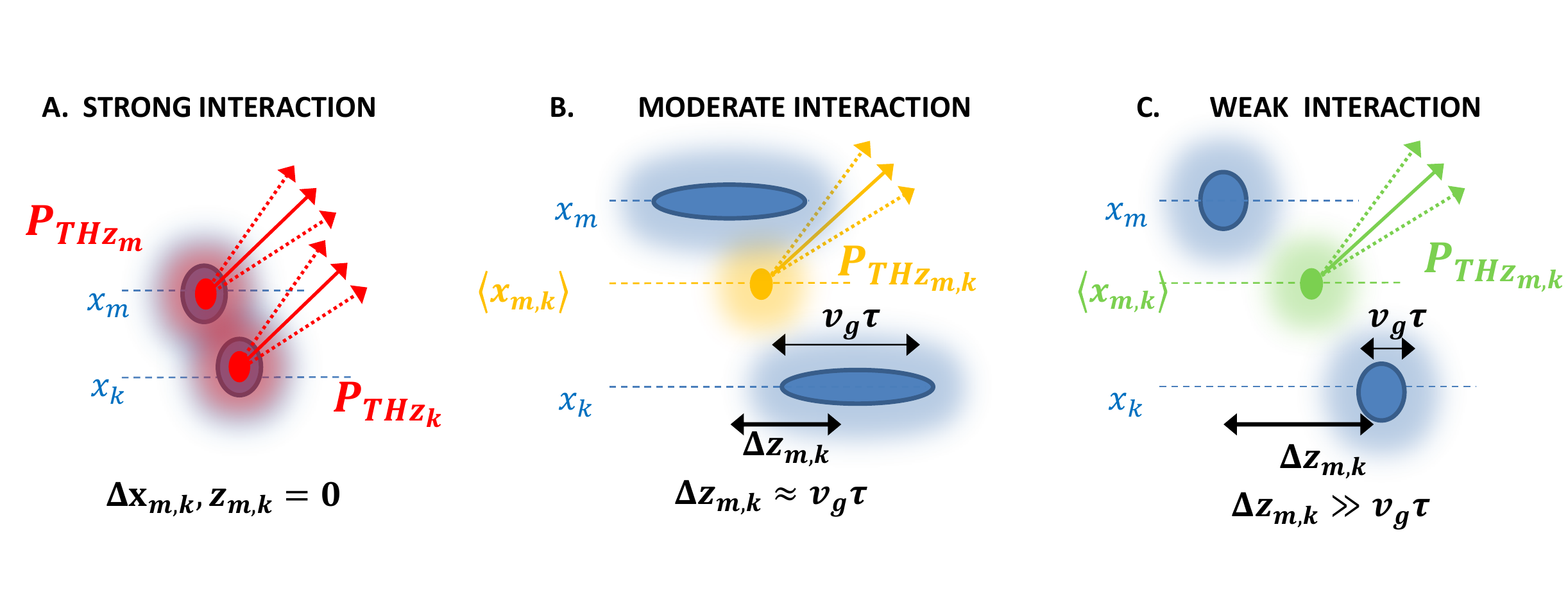}}
\caption{(a) Nonlinear interaction (Eq. (\ref{eq:7_1a})) is strongest within beamlets. (b) Interactions are moderately strong when longitudinal spacing $\Delta z_{p,q}\approx v_g\tau$. (c) Weakest interactions between beamlets occur when transverse and longitudinal spacing is large.}
\label{fig3}
\end{figure}

Cases with strong, moderate, and weak interaction strengths are depicted in Figs. \ref{fig3}(a)-\ref{fig3}(c). Here, the interaction strength decreases as transverse and longitudinal separations grow relative to the beamlet size, $\Delta x$ (i.e. $\Delta x/\sigma>1$), and the temporal width, $v_g\tau$ (i.e. $\Delta z/(v_{g}\tau) >1$). For weak interactions, Eq. (\ref{eq:7_1a}) reduces to a form which is purely a superposition of radiation generated by individual beamlets (only $p=q$ terms are considered), 

\begin{multline}
P_{THz}(x,y,z,\Omega) = \varepsilon_0\chi^{(2)}\frac{\sqrt{2\pi}}{\tau}\frac{\sigma}{\sigma(z)}e^{-\frac{2y^2}{w_y^2}}e^{-\frac{\Omega^2\tau_1^2}{8}}e^{-j\Omega n_gzc^{-1}} \times \\
\sum_{q}E_q^2e^{j\Omega n_gz_{q}c^{-1}}e^{-\frac{2(x- x_{q})^2}{\sigma^2(z)}}e^{-j\frac{v_g^{-1}\Omega(x-x_{q})^2}{2R(z)}}.\label{eq:8} 
\end{multline}

\begin{figure}
\begin{center}
\scalebox{0.4}[0.4]{\includegraphics{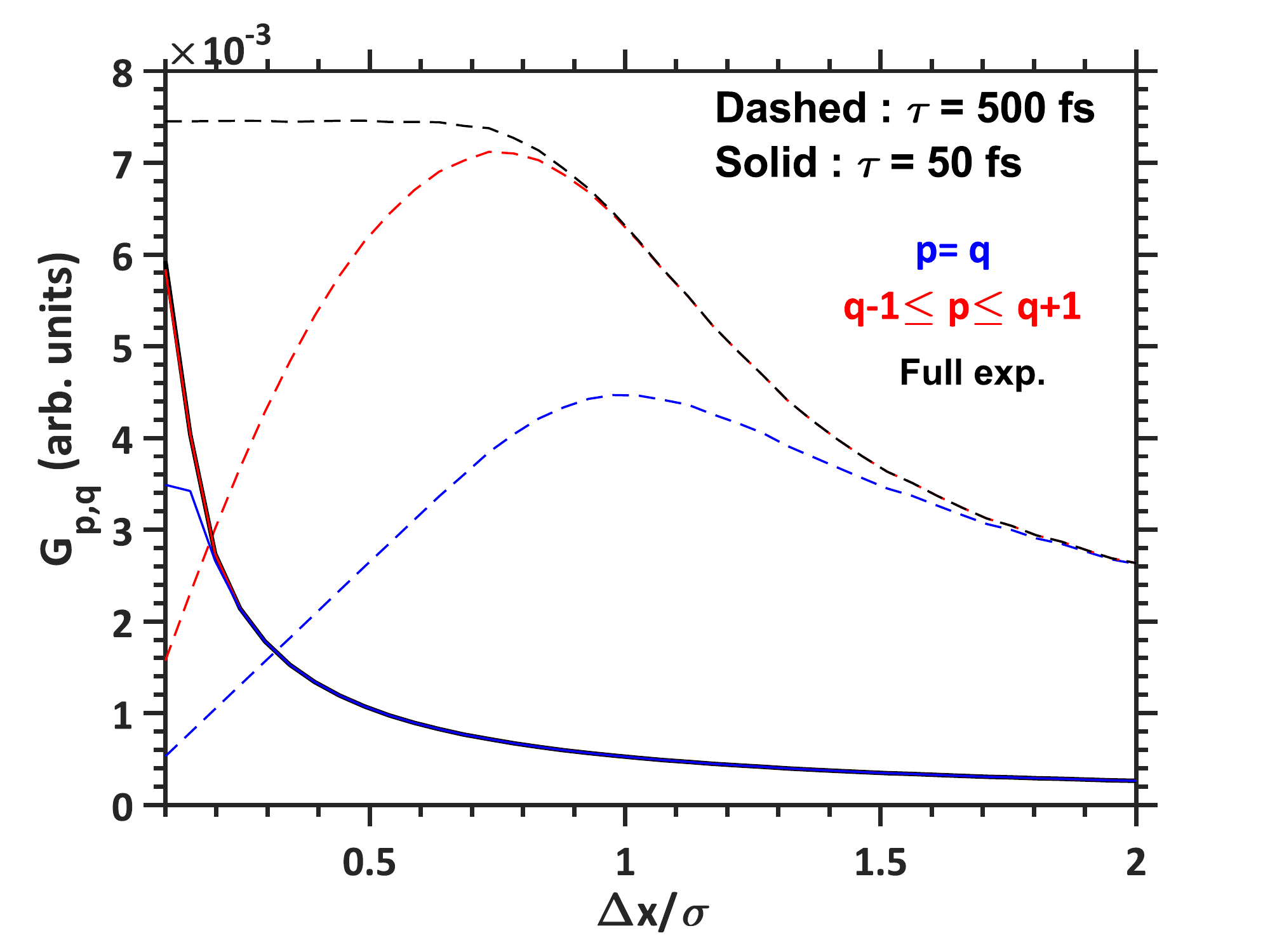}}
\caption{Plots of $G_{p,q}$ for different values of pulse duration and beamlet separation $\Delta x/\sigma$. The peak pump electric field is kept constant for all $\Delta x/\sigma$ and $\tau$. When $v_g\tau$ is large as is the case for $\tau=500$\,fs in lithium niobate, the consideration of interaction between beamlets of the type depicted in Fig. \ref{fig3}(b) become important for small $\Delta x/\sigma$. On the other hand, for $\tau=50$\,fs, interaction within beamlets dominate the overall nonlinear polarization due to small values of $v_g\tau$. 
}\label{fig_int}
\end{center}
\end{figure}

In general, all interaction terms must be considered. This is demonstrated in Fig. \ref{fig_int}, where we plot $G_{p,q}$ while keeping the peak electric field of the pump, $E_{max}$, constant. From Eq. (\ref{eq:5a}), $E_{max}$ may be deduced as,

\begin{gather}
E_{max} = \sum_q \frac{2\pi^{1/2}}{\tau}E_qe^{-x_q^2/\sigma^2}e^{-z_q^2/(v_g\tau)^2}.\label{peak_field}
\end{gather} 

\noindent Fig. \ref{fig_int} shows $G_{p,q}$ for various values of $\Delta x/\sigma$, with $\tau= 500$ fs (dashed) and $\tau=50$ fs (solid). The material is assumed to be lithium niobate. Since $\Delta z\approx v_g\tau$ for $\tau=500$ fs, the full expansion over indices $p,q$ (black, dashed) is necessary as the beamlets get closer or when $\Delta x/\sigma<1$. As $\Delta x\approx \sigma $, only nearest neighbour couplings (i.e. $q=p$ or $q= p\pm1$) become important (red, dashed). This region contains interactions of the form depicted in Fig. \ref{fig3}(b) due to the large value of $v_g\tau$ in relation to $\Delta z_{p,q}$. For even larger separations, only interaction within beamlets (blue, dashed) are required as seen by the convergence of the three curves. For $\tau=50$ fs, $v_g\tau$ is small and only interactions within and between neighboring beamlets are important except at very small separations.

The main distinction between beamlet superposition and the case of grating-based tilted pulse fronts (see Eq. (\ref{tau_2_tpf})) is the absence of a spatial dependence of $\tau$ in Eq. (\ref{eq:7_1a}) and (\ref{eq:8}), which indicates relatively homogeneous spectral properties compared to tilted pulse fronts generated by diffraction gratings. 

\subsection{Terahertz spectra}

From the derived nonlinear polarization term, one may obtain a closed-form expression for $E_{THz}(k_x,y,z, \Omega)$ in $(k_x,\Omega)$-space by Fourier decomposition (see appendix for derivation),

\begin{multline}
E_{THz}(k_x,y,z,\Omega)= -\frac{j\Omega^2\chi^{(2)}e^{-\frac{2y^2}{w_y^2}}}{2k_z(\Omega)c^2}\frac{\sigma b^{1/2}}{\sigma(z)\tau}e^{-\frac{\Omega^2\tau_1^2}{8}}\times \\
\sum_{p}\sum_{q}G_{p,q}e^{j\frac{\Omega \langle z_{p,q}\rangle}{v_g}} e^{jk_x\langle x_{p,q}\rangle}e^{-\frac{k_x^2b}{8}}\textbf{D}(\Delta k,z),\label{eq_ekx}
\end{multline}

\begin{gather}
b^{-1}=\bigg[\frac{1}{\sigma^2(z)}+j\frac{\Omega }{4R(z)v_g}\bigg] \label{b_eff},\\
\textbf{D}(\Delta k,z)=\frac{1}{\frac{\alpha}{2\text{cos}\gamma}+j\Delta k}\bigg[e^{-j\Omega z/ v_g}-e^{-\frac{\alpha z}{2\text{cos}\gamma}}e^{-jk_z(\Omega)z}\bigg].\label{phase_match}
\end{gather}

The above expression can be generally utilized to evaluate properties of the generated terahertz pulse by beamlet superposition in the undepleted limit. We describe various features of this equation below.

\subsubsection{Beamlet superposition}

The effect of coherent superposition of beamlets is represented in the complex exponential terms within the summation in Eq. (\ref{eq_ekx}). For $\Delta z= -\Delta x \text{tan}\gamma$, the terms $\Omega \Delta\langle z_{p,q} \rangle /v_{g} + k_x\Delta\langle x_{p,q} \rangle =0$ at $k_x = k(\Omega)\text{sin}\gamma$, leading to constructive interference between all pairs of beamlets. This corresponds to terahertz radiation propagating at an angle $\gamma$ with respect to the pump beamlets. The $e^{-k_x^2b/8}$ term in Eq. (\ref{eq_ekx}) represents the transverse momentum distribution in $k_x$, with an effective complex beam radius given by $b^{1/2}$, where $b$ is defined in Eq. (\ref{b_eff}).

\subsubsection{Phase matching}

In Eq. (\ref{eq_ekx}), $\textbf{D}(\Delta k,z)$ represents the accumulation of terahertz radiation along the $z$-direction and is given by Eq. (\ref{phase_match}). $\textbf{D}(\Delta k,z)$ is maximized for $k_z=\Omega/ v_g$. In the absence of absorption, $\textbf{D}(\Delta k,z)$ reduces to the well-known $\text{sinc}(\Delta kz/2)z$ function, where $\Delta k=k_z(\Omega,k_x)-\Omega /v_g$. 

\begin{figure}
\begin{center}
\scalebox{0.23}[0.23]{\includegraphics{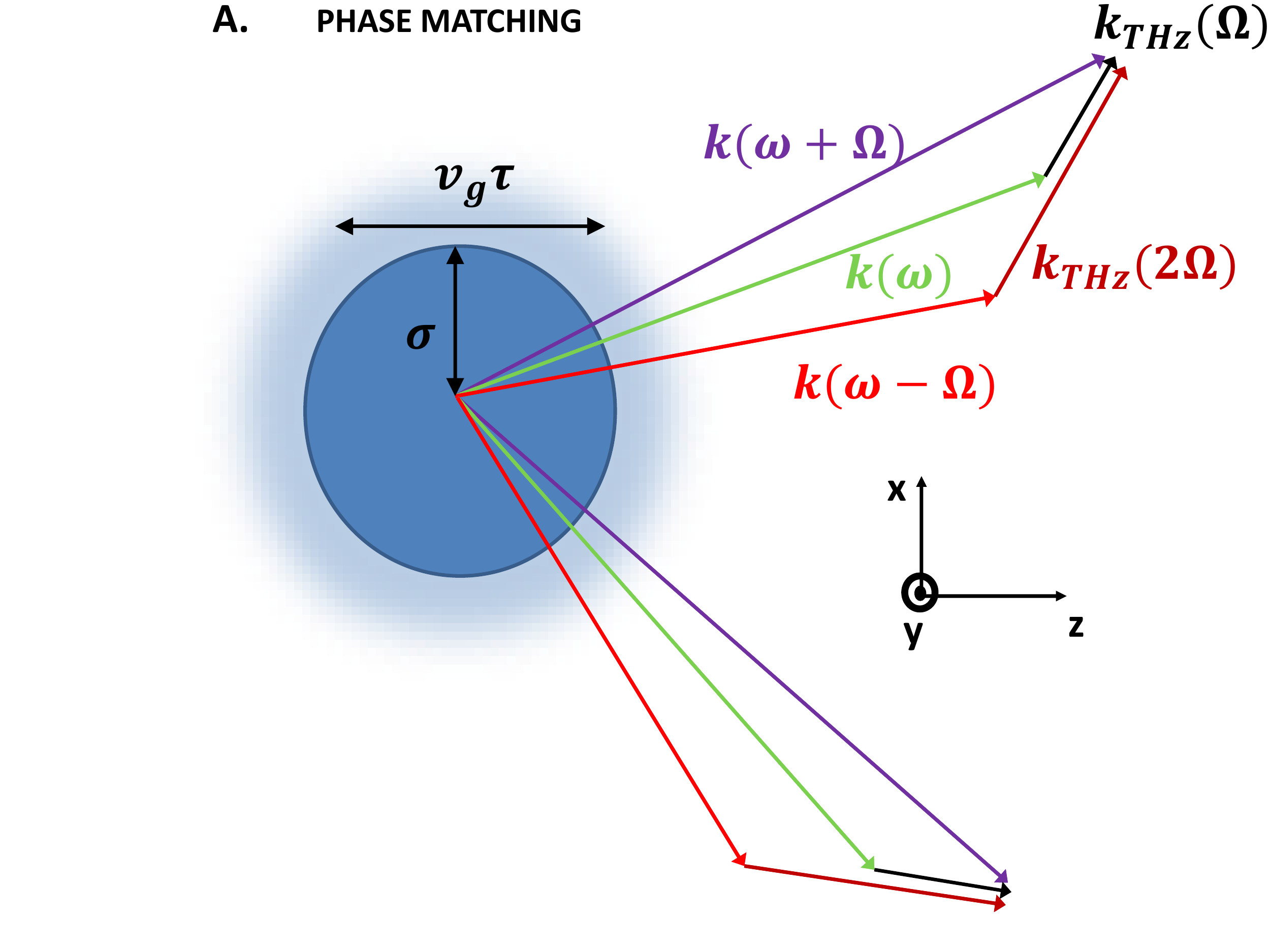}}
\scalebox{0.23}[0.23]{\includegraphics{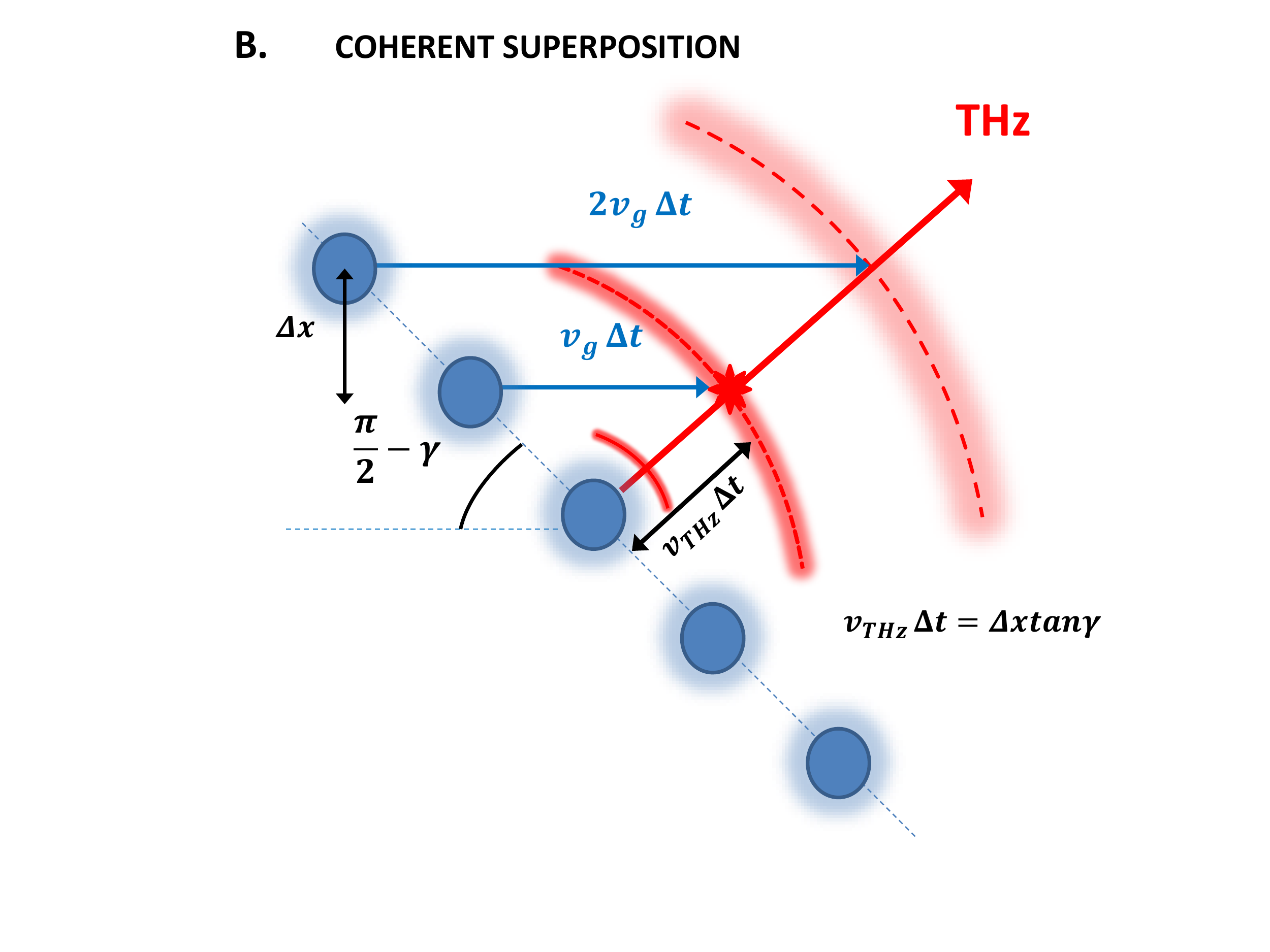}}
\caption{\label{fig5}(a) Larger terahertz frequencies are phase matched at larger values of $k_z$ and hence require greater angular spread or equivalently, smaller beamlet sizes $\sigma$.(b) Terahertz radiation from beamlet $q-1$ only superposes with terahertz generated by beamlet $q$, due to its rapid diffraction. This requires the spacing between beamlets $\Delta x\leq \sigma$, consistent with Eq. (\ref{dx_c}). The diffraction of terahertz radiation and optical beamlets places a lower bound on the value of $\sigma$ }
\end{center}
\end{figure}

\subsubsection{Beam size dependencies}

From Eq. (\ref{eq_ekx}), it is evident that to optimally utilize larger pump bandwidths to generate larger terahertz frequencies, a larger $k_x$-spread is necessary. This is because larger terahertz frequencies will be phase matched at larger values of $k_z$ (see Fig. \ref{fig5}(a)), which necessitates a greater spread in transverse momentum for the  $e^{-k_x^2b/8}$ and $\textbf{D}(\Delta k,z)$ terms in Eq. (\ref{eq_ekx}) to overlap. Therefore, smaller beamlet sizes will be necessary for larger terahertz frequencies. For a desired terahertz wavelength $\lambda_{THz}$, the requisite beamlet size $\sigma$ maybe obtained by setting the $k_x$-distribution in Eq. (\ref{eq_ekx}) to be equal to $e^{-1}$ at $k(\Omega)\text{sin}\gamma$. This yields an upper limit for $\sigma$,

\begin{gather}
\sigma \leq \frac{\lambda_{THz}}{\pi n_{THz}\text{sin}\gamma}.\label{beam_th}
\end{gather}

\subsubsection{Terahertz and beamlet diffraction}

The small values of $\sigma$ in relation to the wavelength of the terahertz radiation causes rapid diffraction. However, if the beamlets are spaced by a distance $\Delta x \leq \sigma$ as posited in Eq. (\ref{dx_c}), the generated terahertz radiation will grow coherently  by superposing with radiation generated by adjacent beamlets. However, they do not coherently interfere with radiation generated by beamlets farther away, as they fall outside the terahertz Rayleigh range as depicted in Fig. \ref{fig5}(b). 

Another salient feature, evident in Eq. (\ref{eq_ekx}), is the longitudinal dependence on distance of the beamlet size $\sigma(z)$ and radius of curvature $R(z)$. The smaller beamlet radius $\sigma$ results in smaller Rayleigh distances in comparison to conventional tilted pulse fronts produced by diffraction gratings, which results in non-negligible diffraction of the pump beamlet. As shall be discussed later, the diffraction of terahertz radiation and pump beamlets place a lower bound on beamlet sizes.

\subsubsection{Closed-form expressions}

While Eq. (\ref{eq_ekx}) is useful in obtaining the distribution of the generated terahertz field in $(\Omega,k_x)$-space, it does not directly lend itself to providing intuition in physical space or temporal domains. Complex functions of $k_x$ prevent obtaining closed-form expressions in the general case. However, closed-form expressions may be derived in the limit of large absorption, i.e. for $\alpha\gg\Delta k$ or when the distribution in $k_x$-space is strongly localized around $\Delta k=0$. For tilted pulse fronts obtained from diffraction gratings, the latter condition is better satisfied in comparison to the case of beamlet superposition. However, qualitative behavior can still be explained by closed-form expressions in this limit.

Criteria for the validity of these approximations and comparisons to exact numerical calculations are established in \cite{ravi2018} for the case of diffraction-grating-based tilted pulse fronts. Similar criteria hold for the beamlet superposition case. 

For quantitative predictions on efficiency and field, one should evaluate Eq. (\ref{eq_ekx}). In the remainder of this section, we illustrate the general physics of this system, using the derived closed-form expressions.

Following the procedure outlined in the appendix, we obtain the following equation for the terahertz spectrum $E_{THz}(\Omega,x,z)$ in ($x,\Omega$)-space,

\begin{multline}
E_{THz}(x,y,z,\Omega)=
-\frac{j\Omega\chi^{(2)}}{\alpha n_{THz}c}\frac{\sqrt{2\pi}}{\tau}\frac{\sigma}{\sigma(z)}e^{-\frac{2y^2}{w_y^2}}e^{-\frac{\Omega^2\tau_1^2}{8}}e^{-j\Omega z /v_{g}}\times\\
\sum_{p}\sum_{q}G_{p,q}e^{j\Omega \langle z_{p,q}\rangle/ v_g}
\bigg[S_{p,q}(x,z,\Omega)-F_{p,q}(x,z,\Omega)\bigg],\label{e_r_z}
\end{multline}
\begin{multline}
F_{p,q}(x,z,\Omega) = \frac{e^{-\frac{\alpha z}{2\text{cos}\gamma}}}{a(z)^{1/2}}
e^{\frac{-2\big(x-\langle x_{p,q}\rangle-z\text{tan}\gamma\big)^2}{\sigma^2(z)|a(z)|^2}}\times \\
e^{-jk(\Omega)\text{sin}\gamma x\left(1-\frac{1}{|a(z)|^2}\right)}
 e^{-\frac{j\Omega}{2R(z)|a(z)|^2v_g}\big(x-\langle x_{p,q}\rangle-z\text{tan}\gamma\big)^2},\label{prop_term}
 \end{multline}
 \begin{gather}
 a(z) = 1-\frac{j4z}{\sigma^2(z)k(\Omega)\text{cos}\gamma}.\label{a_eff}
\end{gather}

\noindent The only new term in Eq. (\ref{e_r_z}) is $F_{p,q}$, which corresponds to the propagating terahertz wave (or far-field term) that is generated by the pump beamlets. This can be compared to $S_{p,q}$, introduced in Eq. (\ref{source_term}), which may be viewed as a source term (or near-field term)  which drives terahertz generation between beamlets $p,q$. From Eq. (\ref{prop_term}), it is seen that $F_{p,q}$ is centered about $x-\langle x_{p,q}\rangle-z\text{tan}\gamma$, indicating that the terahertz wave  propagates at an angle $\gamma$ with respect to the pump. The $e^{-\alpha z/2\text{cos}\gamma}$ factor in Eq. (\ref{prop_term}) indicates attenuation as the terahertz field propagates.

The effect of terahertz beam diffraction is captured by terms with $a(z)$ factors.  The $1/a(z)^{1/2}$ prefactor in Eq. (\ref{prop_term}) indicates a reduction in the strength of the beam. The terahertz beam size increases due to terahertz diffraction by a factor, $|a(z)|^2$. Furthermore, the $1-1/|a(z)|^2$ factors indicate that larger terahertz frequencies diffract less. The rapid diffraction of terahertz beams implies that only terahertz radiation generated by beamlet $q-1$ superposes on top of that produced by beamlet $q$ (see Fig. \ref{fig5}(b)).

\subsection{Temporal profiles}

The spatio-temporal terahertz field profiles, $E_{THz}(x,y,z,t)$, are obtained by inverse Fourier transformation of Eq. (\ref{e_r_z}). However, the presence of frequency-dependent terms in the denominator of Eq. (\ref{e_r_z}) makes this challenging. To eliminate these dependencies while retaining essential features, we set $k(\Omega)\text{cos}\gamma/2\approx\sigma$ in $a(z)$. This approximation is equivalent to making the wavelength of the generated terahertz pulse comparable to the size of an individual beamlet. Following this substitution, the expression for $E_{THz}(x,y,z,t)$ is,

\begin{multline}
E_{THz}(x,y,z,t)=
\frac{-8\pi\chi^{(2)}}{\alpha n_{THz}c\tau\tau_1^3}\frac{\sigma}{\sigma(z)}e^{-\frac{2y^2}{w_y^2}}
\bigg[\sum_{p}\sum_{q}G_{p,q}
\bigg(e^{-\frac{2(x-\langle x_{p,q}\rangle)^2}{\sigma^2(z)}}t'e^{-\frac{2t'^2}{\tau_1^2}}\\
-e^{-\frac{\alpha z}{2\text{cos}\gamma}}\frac{\text{cos}[\frac{1}{2}\text{tan}^{-1}\frac{2z}{\sigma}]}{\sqrt{1+\frac{4z^2}{\sigma^2(z)}}}
e^{\frac{-2\big(x-\langle x_{p,q}\rangle-z\text{tan}\gamma \big)^2}{\sigma^2(z)\big[1+\frac{4z^2}{\sigma^2(z)}\big]}}t''e^{-\frac{2t"^2}{\tau_1^2}}\bigg)\bigg],\label{e_t}
\end{multline}

\begin{gather}
t' = t-\frac{z-\langle z_{p,q}\rangle}{v_g}+\frac{(x-\langle x_{p,q}\rangle)^2}{2R(z)v_g},\label{t_1}\\
t'' = t-\frac{z-\langle z_{p,q}\rangle}{v_g}- \frac{4z^2}{4z^2+\sigma^2(z)}\frac{\text{sin}\gamma(x-\langle x_{p,q}\rangle)}{v_{THz}}+\frac{\big(x-\langle x_{p,q}\rangle-z\text{tan}\gamma \big)^2}{2R(z)\big[1+\frac{4z^2}{\sigma^2(z)}\big]v_g}.\label{t_2}
\end{gather}

\begin{figure}
\begin{center}
\scalebox{0.33}[0.33]{\includegraphics{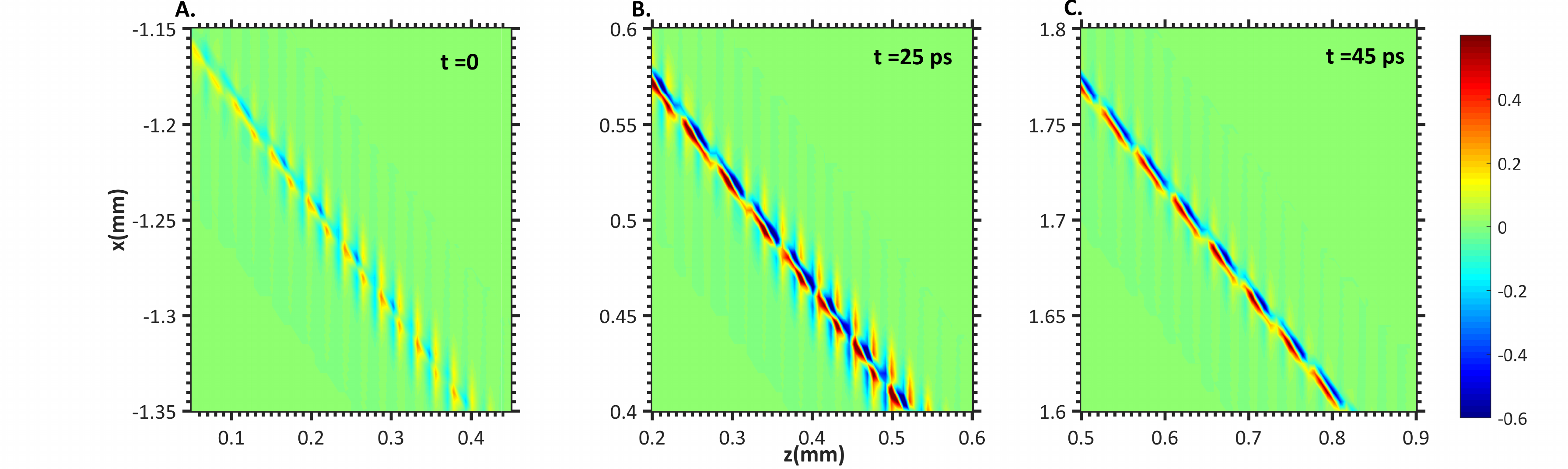}}
\caption{\label{fig12}Temporal evolution of terahertz transients obtained from Eqs. (\ref{e_t})-(\ref{t_2}) are plotted for $\tau=50\,$fs, $\sigma=25\,\mu$m and $\Delta x=25\,\mu$m.(a) A discrete set of beamlets produces a series of terahertz transients. (b) These grow in amplitude and diffract. (c) Gradually, a smooth tilted pulse fronts is formed. Diffraction at the exit surface of the crystal would further wash out the discreteness of the tilted pulse front.}
\end{center}
\end{figure}

From Eq. (\ref{e_t}), there are a series of terahertz beamlets centered about the mean positions of the pump beamlets $p,q$, each with $1/e^2$ beam radius $\sigma(z)/\sqrt{2}$ and pulse duration $\tau_1(z)/\sqrt{2}$. The $\sqrt{2}$ factor arises from the nature of second-order interactions. The $z$ dependence of the terahertz pulse duration arises from the balance between material dispersion and any external group-delay dispersion experienced by the pump pulse. Each terahertz beamlet is a sum of a source (near-field) term and a propagating (far-field) term, analogous to Eq. (\ref{e_r_z}) and \cite{bakunov2008,bakunov2011,bakunov2014,ravi2018}. The temporal variation of the terahertz waveform is given by $\sim te^{-2t^2/\tau^2}$, where $t$ is the time variable defined in an appropriate frame of reference. For instance, Eq. (\ref{t_1}) represents the time variable of the near-field term, $t'$. On the other hand, Eq. (\ref{t_2}) represents the time variable of the near-field term, $t''$. The near-field term is roughly $z$ propagating (for large $R(z)$) as can be seen by inspecting Eq. (\ref{t_1}). Equation (\ref{t_1}) also indicates that points along the radius of curvature are of equal phase.

The second term of Eq. (\ref{t_2}) implies that close to the source location (i.e. $z\approx 0$), the field is virtually $z$ propagating because the second term vanishes. When $z\sim\sigma$, diffraction effects produce a propagating terahertz wave predominantly moving at an angle $\gamma$ with respect to the pump beamlets. As the terahertz wave propagates, its radius of curvature continues to increase as evident from the final term in Eq. (\ref{t_2}).

In Fig. \ref{fig12}, we plot snapshots of the terahertz electric field normalized to the peak value obtained from Eq. (\ref{e_t}) for lithium niobate with $\gamma = 62^\circ, n_g=2.25 , n_{THz}= 4.73$ at T= 100\,K. Other parameters are tabulated in Table \ref{param_list}. In Fig. \ref{fig12}(a), the field is weak and both near-field and far-field terms contain a plane of constant phase along $t-z/v_g$ (if one ignores the effects of the radius of curvature). As the field propagates, its strength increases due to phase-matched terahertz generation. In addition, the second term in Eq. (\ref{t_2}) begins to grow due to terahertz diffraction. The terahertz field gradually becomes aligned along the desired tilt plane, i.e. $t-x\text{sin}\gamma/v_{THz} -z/v_g$ as can be seen in Fig. \ref{fig12}(c). Discreteness in the tilted pulse front further washes out with propagation, and the end result is a terahertz plane wave.

\section{Results and discussion}
\begin{table*}[ht]
\captionsetup{labelfont=bf}
\centering
\caption{\label{param_list} \textbf{List of parameters used in calculations.}}
\begin{tabular}{l c r}
\hline
Parameter & Symbol & Value \\
\hline
Second order nonlinear coefficient & $\chi^{(2)}$ & 336\,pm/V  \\

Central pump wavelength &$\lambda_0$   & 1.03\,$\mu$m   \\

Optical phase index   & $n_{NIR}$ & 2.17  \\

Optical group index   & $n_g$ & 2.25 \\

Terahertz phase index & $n_{THz}$& 4.75 (T = 100\,K)\\
&& 4.95 (T = 300\,K)\\

GVD-AD & $k_T$ & $-1\times 10^{-23}$\,s$^2$/m \\

Full Terahertz index dispersion & $n_{THZ}(\Omega)$ & \cite{palfalvi2005}\\
Full Terahertz absorption dispersion & $\alpha(\Omega)$&\cite{fulop2011}\\
\hline
\end{tabular}
\end{table*}

\subsection{Comparisons to tilted pulse fronts from diffraction gratings}
In this section, we present results that demonstrate the advantages of using a superposition of beamlets as opposed to tilted pulse fronts obtained from diffraction gratings. At the outset, we first present key equations for terahertz radiation generated using DG-TPF's that were derived in a recent analysis \cite{ravi2018}. These account for various spatio-temporal distortions that characterize DG-TPF's.

The  counterpart of Eq. (\ref{eq_ekx}) for DG-TPFs is, 

\begin{gather}
E_{THz}(\Omega,k_x,z)|_{TPF} =  \frac{-j\Omega^2w_1\chi^{(2)}(z)E_0^2}{2k_z(\Omega)c^2\tau}e^{-\frac{\Omega^2\tau^2}{8}}e^{-\frac{(k_x-k(\Omega)\beta v_g)^2w_1^2}{8}}e^{-jk(\Omega)\text{sin}\gamma x}\textbf{D}(\Delta k, z),\label{tpf_kx}\\
w_1=w\sqrt{\frac{1}{1+\frac{k_T^2\Omega^2w^2}{\tau^2}}}.\label{w_eff}
\end{gather}

The above equation is generally valid and can be used for quantitative comparisons to the case of beamlet superposition. In comparing Eq. (\ref{eq_ekx}) to Eq. (\ref{tpf_kx}), notice that the distribution in transverse momentum $k_x$ is centered about $k(\Omega)\beta v_g$ and has a much narrower spread in $k_x$ space as $w$ is much larger than $\sigma$. Additionally, there is a reduction in the amplitude of $E_{THZ}(\Omega,k_x,z)$ by a factor of proportional to $(1+k_T\Omega^2w^2/\tau^2)^{-1/2}$ as captured by the effective beam radius term $w_1$ in Eq. \eqref{w_eff}. This shall be proven in a subsequent section to most negatively affect terahertz generation produced by DG-TPF's.

Proceeding along similar lines to that of the beamlet superposition case, one obtains the following closed-form expression for terahertz transients generated by DG-TPF's,

\begin{gather}
E_{THz}(t,x,z) = \frac{-8\pi\chi^{(2)}E_0^2}{\alpha n_{THz}c\tau}
\bigg[\frac{1}{\tau_{1,x}^3}e^{-2x^2w_0^{-2}}t'e^{-\frac{2t'^2}{\tau_{1,x}^2}}
-\frac{1}{\tau_{2,x}^3}e^{-\frac{\alpha z}{2\text{cos}\gamma}}e^{-2(x-z\text{tan}\gamma)^2w_0^{-2}}t''e^{-\frac{2t''^2}{\tau_{2,x}^2}}\bigg],\label{e_t_x_1}\\
\tau_{1,x} =\tau\left[1+\frac{16x^2k_T^2}{\tau^4}\right]^{1/2}~, ~
\tau_{2,x} =\tau\left[1+\frac{16(x-z\text{tan}\gamma)^2k_T^2}{\tau^4}\right]^{1/2},\label{tau_2_tpf}\\
t' = t-\frac{x\text{sin}\gamma}{v_{THz}} -\frac{z}{v_{g}}~,~
t'' = t-\frac{x\text{sin}\gamma}{v_{THz}} -\frac{z}{v_{g}}.\label{t_2_tpf}
\end{gather}

Similar to Eq. (\ref{e_t}), two principal terms corresponding to the source (or near-field) term and propagating (or far-field) term are present. The key difference compared to Eq. (\ref{e_t}) is the presence of a transverse spatial dependence on $\tau_{1,x}$ and $\tau_{2,x}$. $\tau_{1,x},\tau_{2,x}$ produce a terahertz spectrum that is spatially chirped by an amount that is commensurate to GVD-AD ($k_T$) and bandwidth $\tau^{-1}$. From Eq. (\ref{t_2_tpf}), it is evident that both source and propagating terms are constant in phase along the line $ x+z\text{tan}\gamma$. This is different from the case of beamlet superposition, where the field gradually forms the desired tilted pulse front.

\begin{figure}
\begin{center}
\scalebox{0.33}[0.33]{\includegraphics{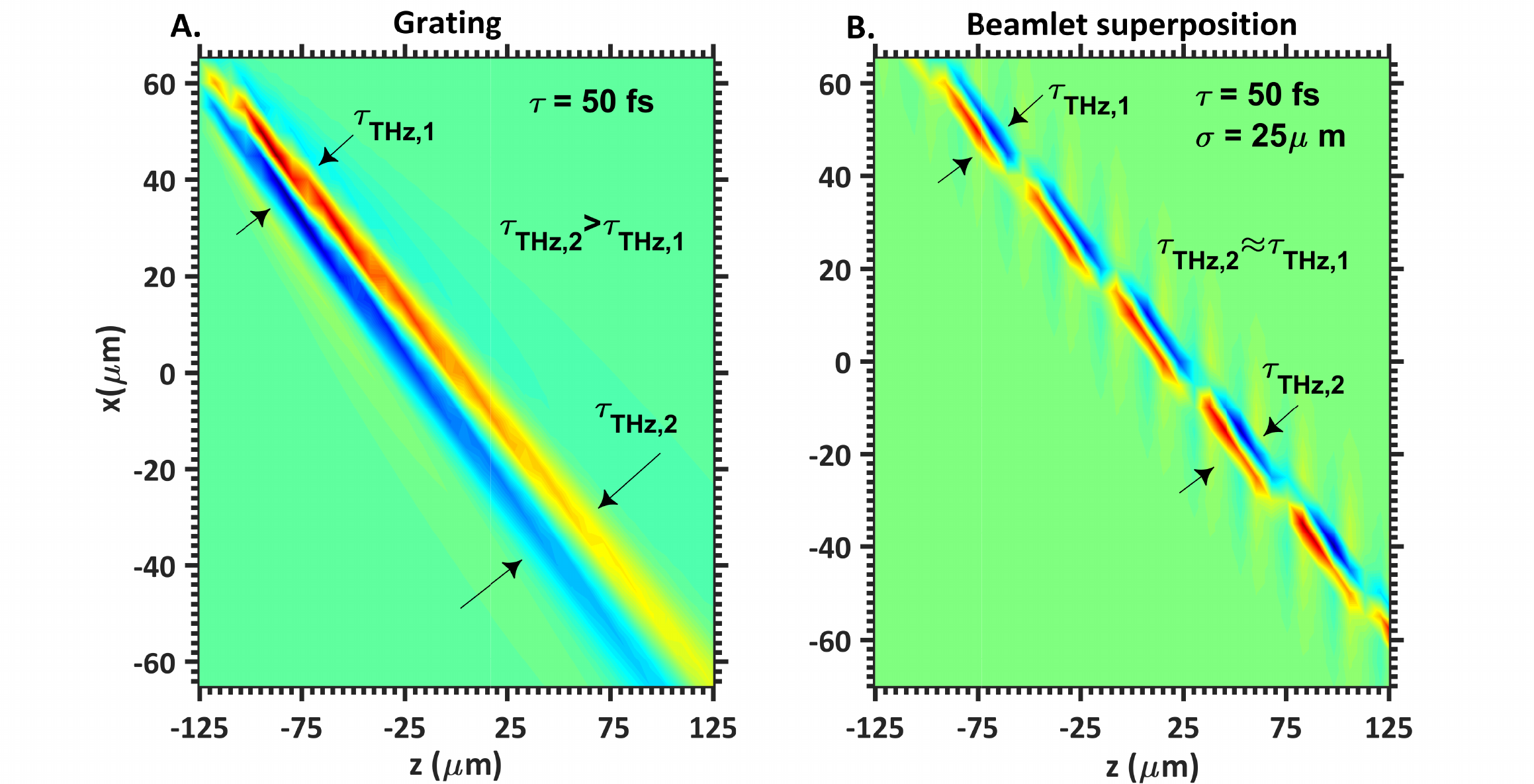}}
\caption{\label{fig13} (a) Terahertz transients generated by tilted pulse fronts produced by DG-TPF's using a pulse with properties $\tau=50$\,fs and $w=2$\,mm. The transient is tilted along the line $x+z\text{tan}\gamma$, where $\gamma$ is the tilt angle and exhibits a drastic variation in pulse duration along this line.(b) A transient generated by a superposition of beamlets with radius $\sigma =25\,\mu $m, spaced by $\Delta x=12.5\,\mu$m, exhibits little transverse variation in terahertz duration/frequency.}
\end{center}
\end{figure}

To illustrate the key differences in terahertz transients formed by beamlet superposition and DG-TPF's, we plot a representative snapshot of the terahertz field for each case with a pump pulse duration of $\tau=50$\,fs in Fig. \ref{fig13} and a tilt angle of $\gamma = 62^\circ$, corresponding to lithium niobate. The assumed GVD-AD value is $k_T=-1\times 10^{-23}\,\text{s}^2\text{m}^{-1}$ \cite{ravi14}. The beamlet size was $\sigma=25\,\mu$m, with beamlet separation $\Delta x=12.5\,\mu$m.

In Fig. \ref{fig13}(a), the variation of terahertz duration or frequency along the tilt-plane is clearly evident for the case of terahertz radiation generated by DG-TPF's. In contrast, terahertz radiation generated by beamlet superposition is characterized by a uniform pulse duration across the tilt-plane .

\subsection{Efficiency and spectra}

Having illustrated a salient difference between terahertz generation by DG-TPFs's and beamlet superposition, we now identify situations when beamlet superposition generates terahertz radiation more efficiently compared to tilted pulse fronts produced by diffraction gratings. All calculations in this section are performed without approximations using the expressions from Eq. (\ref{eq_ekx}) for beamlet superposition and Eq. (\ref{tpf_kx}) for DG-TPF's. The role of optical setups which may impart additional dispersion/aberrations have not been considered in this analysis. However, this is not expected to alter the conclusions of this work. In fact, for beamlet superposition produced by metallic structures such as the echelon \cite{ofori2016thz}, dispersive properties are superior to transmissive optics used in the case of diffraction-grating-based tilted pulse fronts. We use parameters from Table \ref{param_list} for the calculations.

\begin{figure}
\begin{center}
\scalebox{0.35}[0.35]{\includegraphics{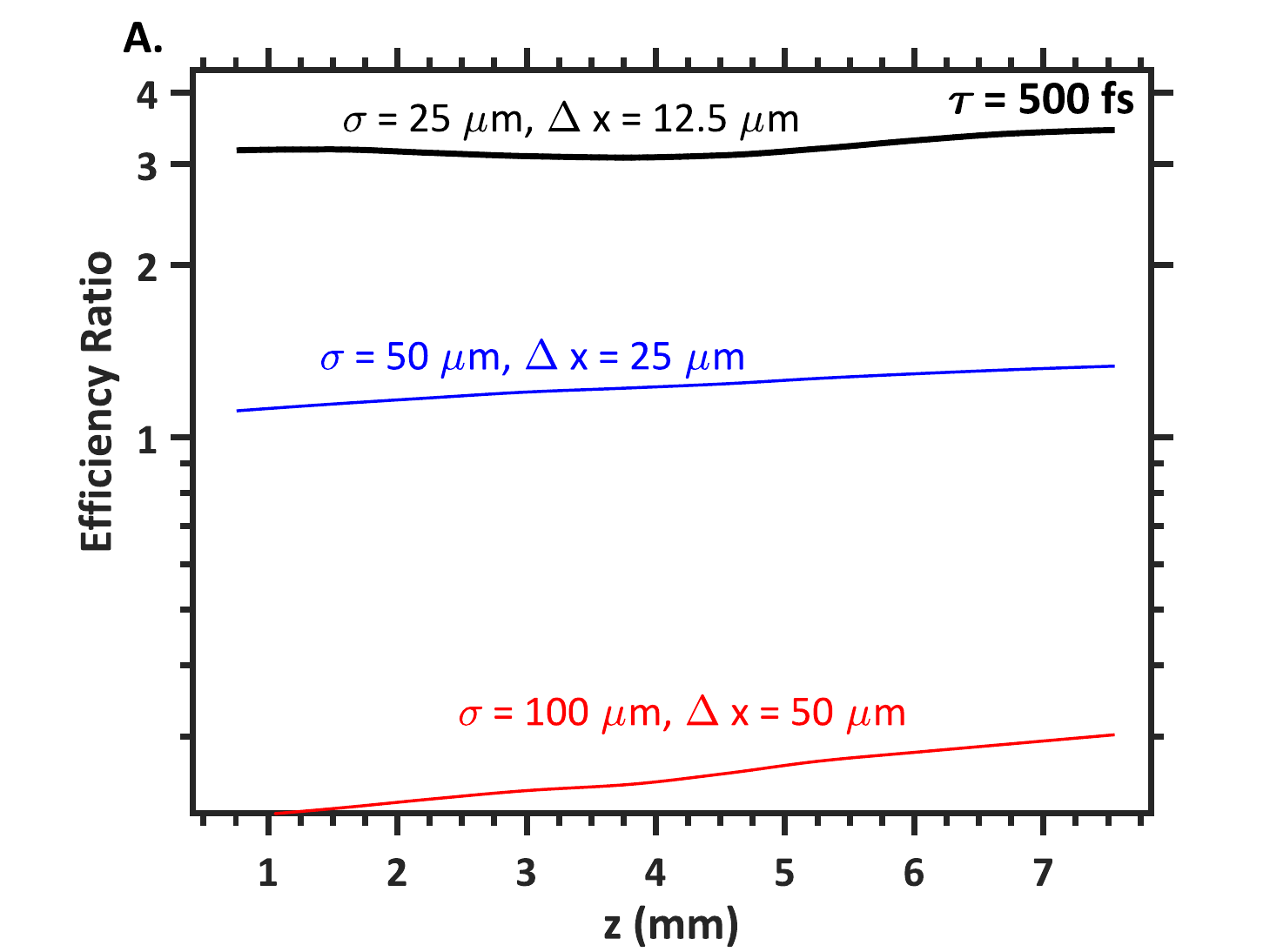}}
\scalebox{0.35}[0.34]{\includegraphics{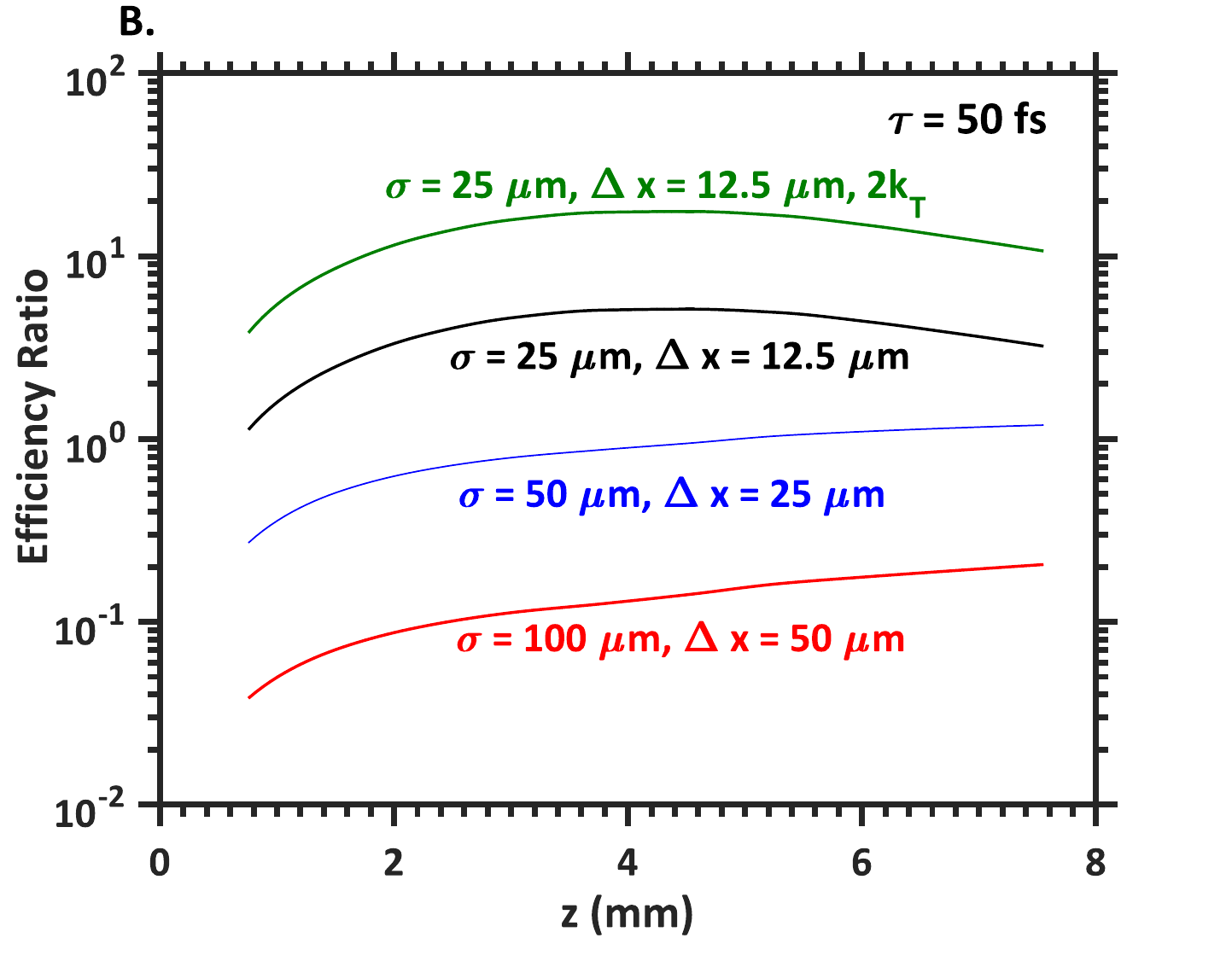}}\\
\hspace{0.25cm}\scalebox{0.35}[0.35]{\includegraphics{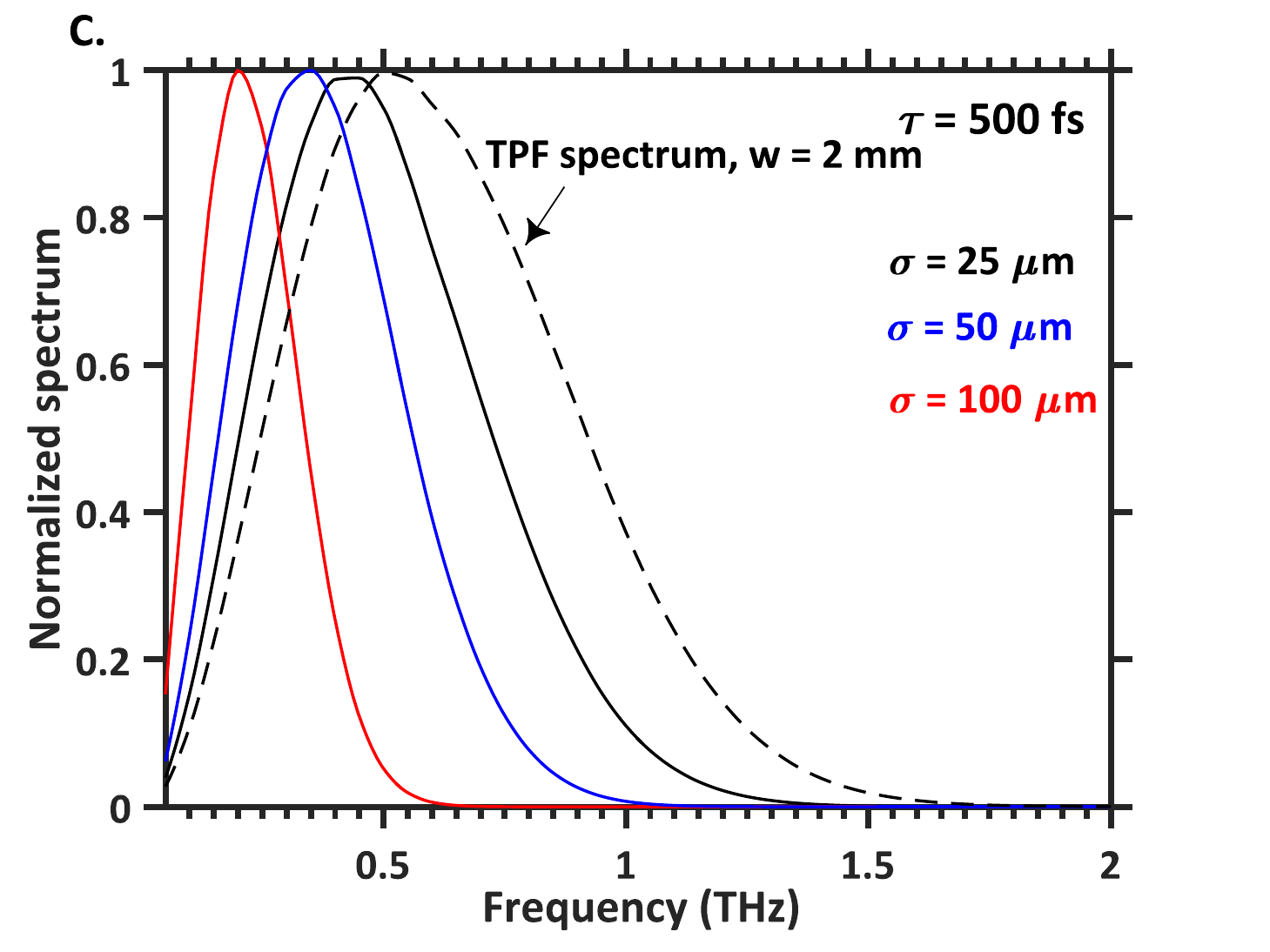}} 
\hspace{-0.25cm}\scalebox{0.35}[0.35]{\includegraphics{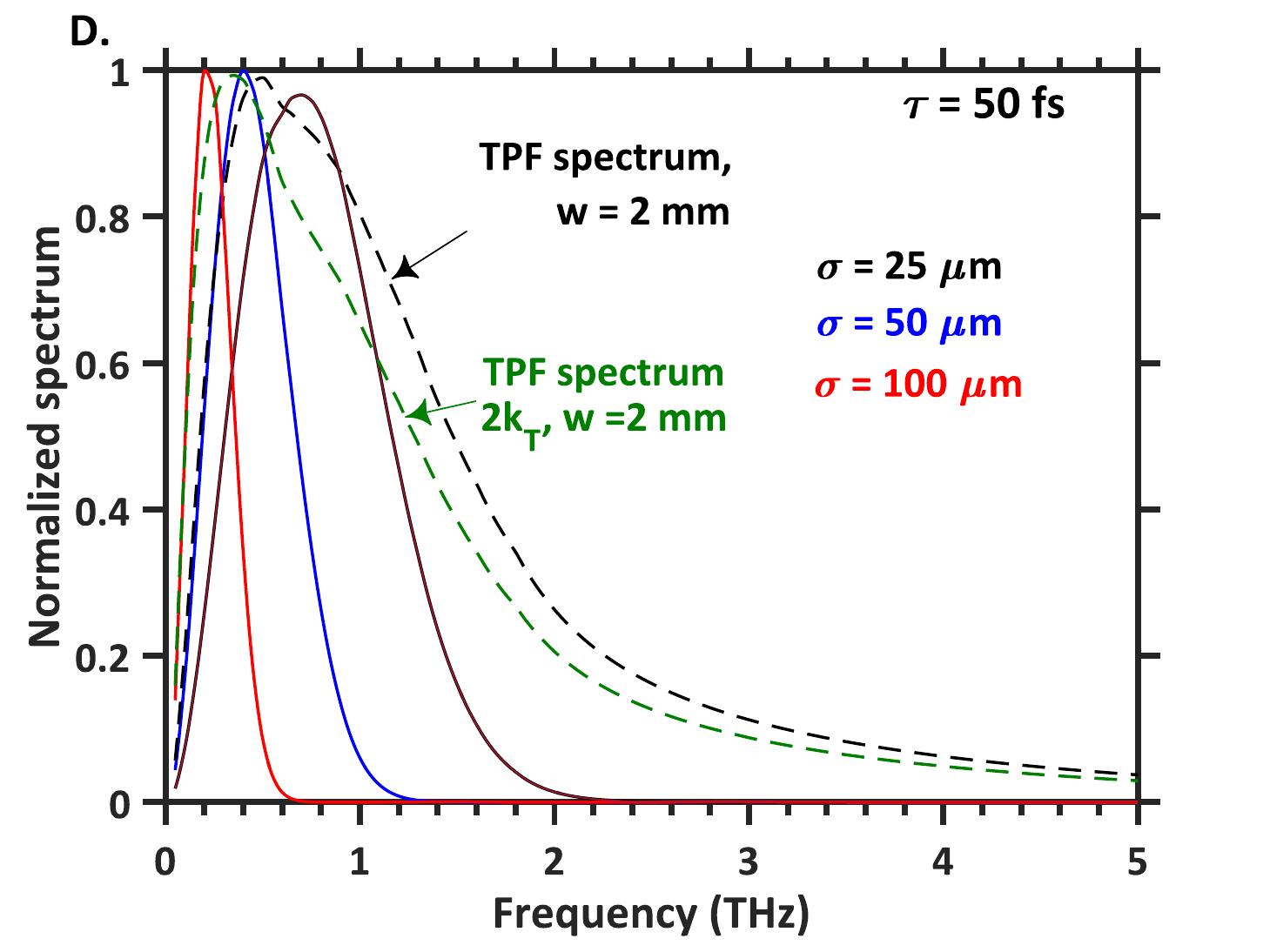}}
\caption{\label{fig8}(a) Ratios of terahertz generation efficiency of beamlet superpostion to that produced by DG-TPF's for $\tau = 500$\,fs. As beamlet sizes $\sigma$ get small enough to supply the necessary bandwidth, terahertz generation efficiencies for beamlet superposition become larger compared to DG-TPF's. (b) Efficiency ratios for $\tau=50$\,fs: The threshold beamlet size is smaller since a larger transverse momentum spread is required to utilize the bandwidth of the incident pump pulse. The ratios increase with larger GVD-AD, due to deterioration of terahertz generation by DG-TPF's. (c) Terahertz spectra for various cases for $\tau=500$\,fs : The reducing beamlet size produces an increase in terahertz frequency. (d) Terahertz spectra for the $\tau =50$\,fs case. Beamlet superposition produces higher frequencies in relation to gratings, with particularly improved performance for larger GVD-AD values.}
\end{center}
\end{figure}

In Fig. \ref{fig8}(a), we plot the conversion efficiency ratios $\eta_{S}/\eta_{DG-TPF}$ of beamlet superposition relative to diffraction gratings, for various beamlet sizes $\sigma$ and a pump duration $\tau=500$ fs. A crystal temperature of $T=100\,$K is assumed. Conversion efficiencies are calculated based on the formulation in the appendix.  In accordance with the threshold condition for beamlet separation outlined in Eq. (\ref{dx_c}), we set $\Delta x= \sigma/2$. It is worth reiterating that the condition in Eq. (\ref{dx_c}) was obtained under the assumption of Gaussian spatial profiles for beamlets. However, in cases such as in the echelon \cite{ofori2016thz}, the beamlets are closer to having top-hat profiles. This would potentially alleviate the proximity requirement for beamlets.

The beam radius for the DG-TPF case is fixed at $w=2$ mm and the total number of beamlets is varied with $\Delta x$ as $N=2w/\Delta x$. The peak pump electric field for DG-TPF ($\sqrt{2\pi}E_0/\tau$) and beamlet superposition (Eq. (\ref{peak_field})) cases are maintained constant. We rely on efficiency ratios rather than absolute values since absolute efficiency values are only meaningful when depleted calculations (i.e cascading effects) are performed.

It can be seen in Fig. \ref{fig8}(a) that the efficiency of terahertz radiation generated by beamlet superposition is smaller than that obtained from gratings for $\sigma=100\,\mu m$ (red). This is because the radius $\sigma$ is too large to contain a sufficient spread in transverse momentum to effectively utilize the pump bandwidth. Only for $\sigma=50\,\mu m$ (blue), is parity in performance obtained. 

As $\sigma$ is reduced to $25$ $\mu m$, the efficiency ratio significantly exceeds unity. In contrast to the cases of $\sigma=100,50\,\mu$m, which exhibit monotonic growth of efficiency over propagation length $z$, for $\sigma= 25\mu m$, terahertz diffraction begins to play a slight role. A minimum in efficiency is observed close to $z=4$\,mm, which is where the beam waist $z_0$ is assumed to be located. 

In Fig. \ref{fig8}(b), efficiency ratios are plotted for $\tau=50\,$fs. In this case, the larger bandwidth necessitates a smaller beamlet size to achieve efficiency parity. The reason for this is that the larger bandwidth contributes to higher terahertz frequency components compared to $\tau = 500\,$fs in the DG-TPF case. This can also be understood dy re-writing Eq. (\ref{beam_th}) as $\sigma \leq c\tau(n_{THz}\text{sin}\gamma)^{-1}$ since the optimal terahertz angular frequency $\Omega_{THz}\approx 2\tau^{-1}$ (which may be obtained by maximizing $\Omega^2e^{-\Omega^2\tau^2/4}$). This means that a shorter pump duration would require a smaller beamlet size to generate terahertz radiation efficiently.

 This is illustrated by the black dashed curves in Figs. \ref{fig8}(c)and \ref{fig8}(d).  However, the beamlet size is not small enough to produce the same high frequency components as can be seen by comparing the terahertz spectra produced by the $\sigma=50\,\mu$m (blue, Fig. \ref{fig8}(d)) and DG-TPF case (black dashed, Fig. \ref{fig8}(d)). However, as the beamlet size is reduced to $25\,\mu$m (black, Fig. \ref{fig8}(b)), the efficiency ratio is much larger compared to that when $\tau =500\,$ fs as the effect of GVD-AD is more adverse for larger bandwidths. Larger values of GVD-AD deteriorate the performance of DG-TPFs's even further (green, Fig. \ref{fig8}(b)), because larger $k_T$ values push the generated terahertz frequency by DG-TPF's to a lower value (green dashed, Fig. \ref{fig8}(d)). Furthermore, there is little effect of terahertz diffraction for the $\tau = 50\,$ fs case as a larger terahertz frequency is generated for the same beamlet size of $25\,\mu$m in the case of the $\tau=50\,$fs pump pulse (black, Fig. \ref{fig8}(d)) compared to the $\tau= 500\,$fs case (black, Fig. \ref{fig8}(c)). The reduction in efficiency exhibited by $\Delta x=12.5\,\mu$m in Fig. \ref{fig8}(b) for longer values of $z$ is due to increased absorption at larger terahertz frequencies. 

\begin{figure}
\begin{center}
\scalebox{0.35}[0.35]{\includegraphics{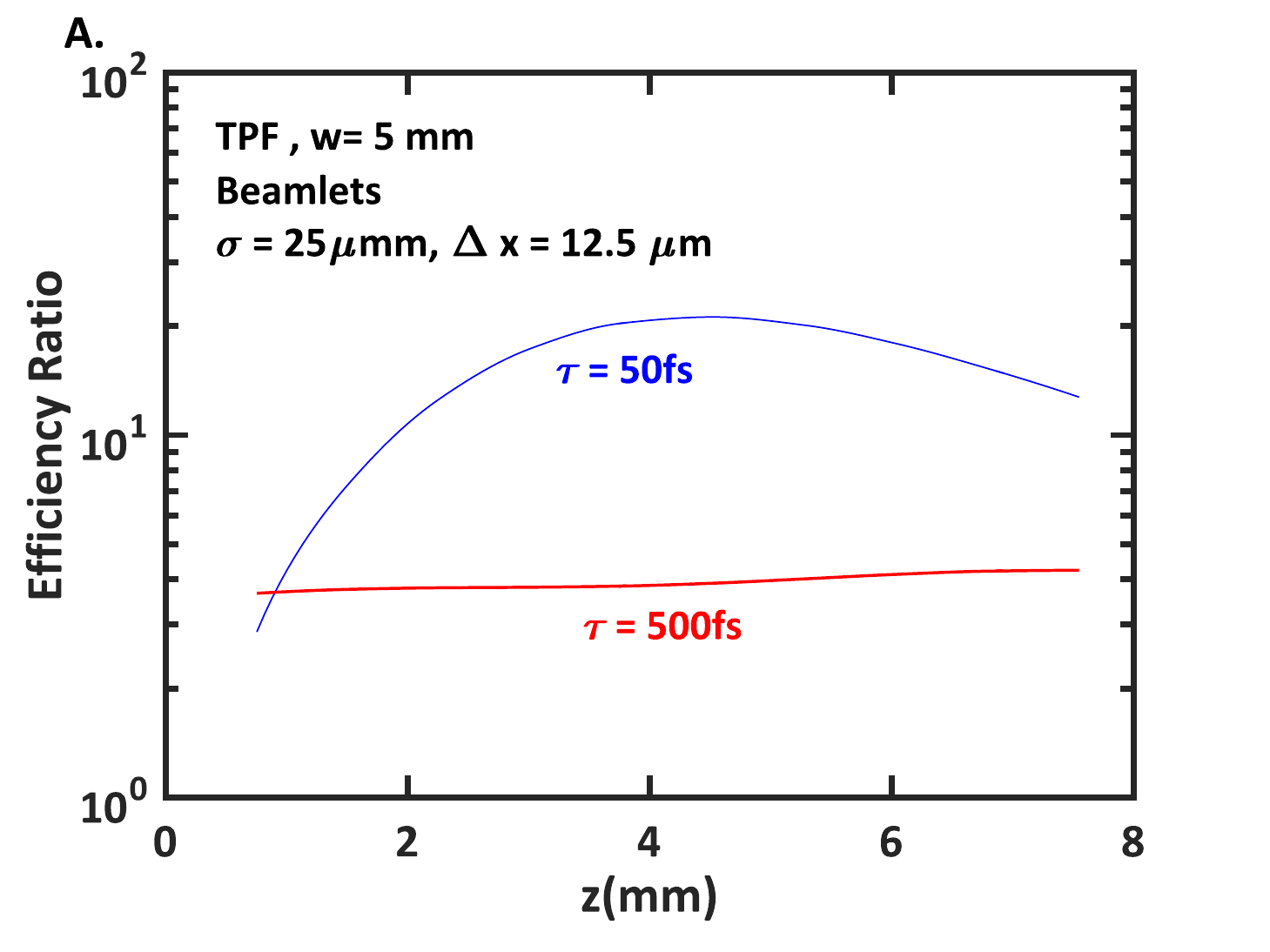}}
\scalebox{0.175}[0.175]{\includegraphics{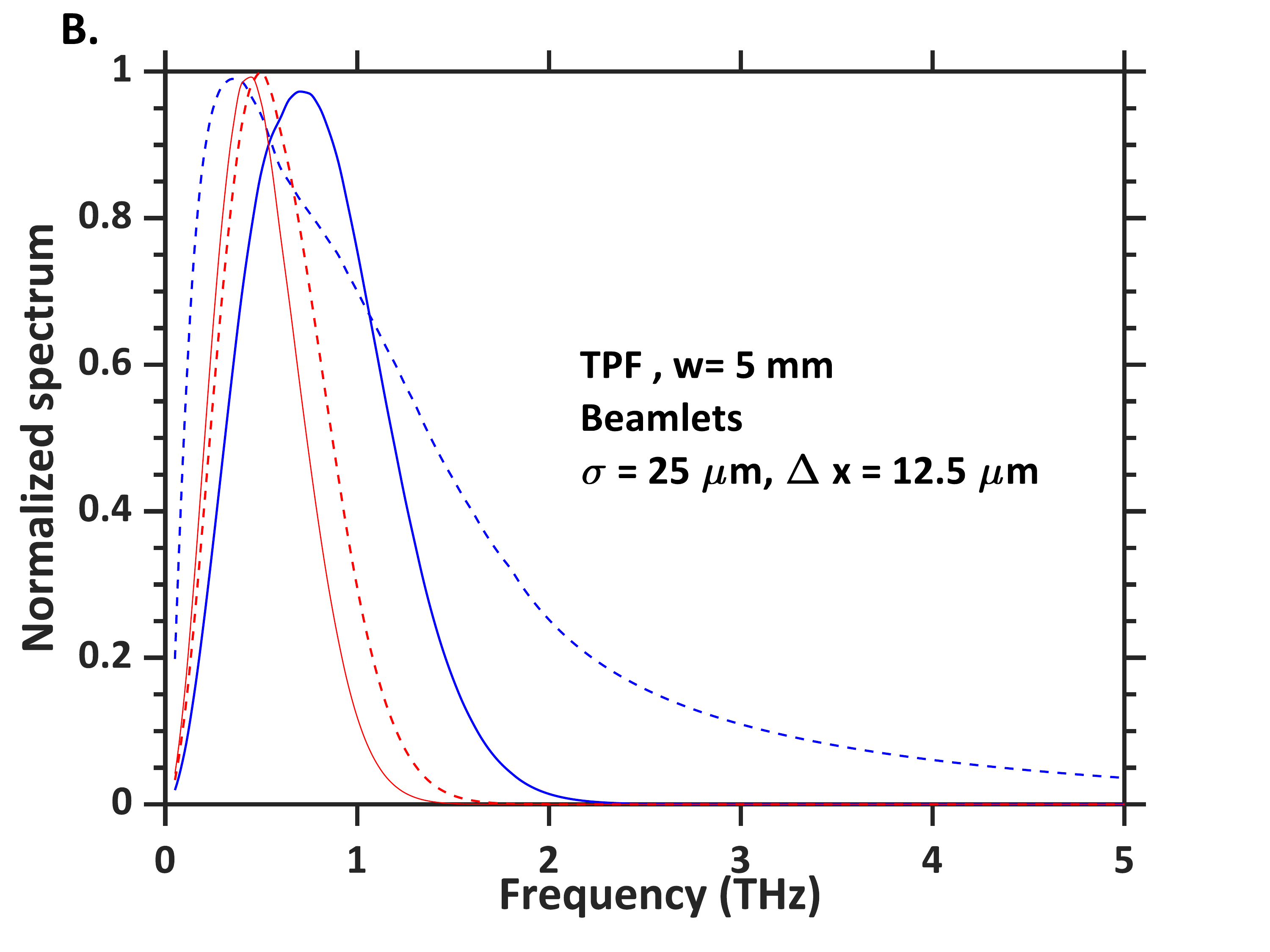}}\caption{\label{fig11}Effect of beam radius (a) Efficiency ratios for a larger total beam radius $w= 5\,$mm. The performance of $\tau= 50\,$fs for beamlet superposition is even further enhanced due to a greater impact of GVD-AD for larger beam sizes in the grating case. (b) Terahertz frequencies for larger beam sizes are smaller for the grating case, while beamlet superposition shows scalable performance.}
\end{center}
\end{figure}

Beamlet superposition is also more scalable with beam size. Figure \ref{fig11} plots the efficiency ratios for $w= 5\,$mm for $\tau = 50\,$fs and $\tau = 500\,$fs. In this case, the efficiency ratios are even larger compared to the corresponding situations for $w=2\,$mm. While the effect is less pronounced for $\tau= 500\,$fs, it is significant for $\tau = 50\,$fs. This is because the effect of GVD-AD is more adverse for larger bandwidths (see Eqs. (\ref{tpf_kx})-(\ref{w_eff}) ). The average frequency also reduces (blue dashed, Fig.\ref{fig11}(b)) compared to the $w=2\,$mm case (black dashed, Fig.\ref{fig8}(d)). Taken together, Fig. \ref{fig8} and Fig. \ref{fig11} indicate that that with the right beamlet size, conversion efficiencies of the beamlet superposition case can significantly exceed those from DG-TPF's particularly for larger bandwidths and total beam sizes. 

\subsection{Analytic proof of efficiency enhancement}

Differences in the conversion efficiency of terahertz radiation generated by DG-TPF's and beamlet superposition may be discerned by approximating the $\textbf{D}(\Delta k,z)$ terms in Eq. (\ref{tpf_kx}) and Eq. (\ref{eq_ekx}) respectively by a Dirac-Delta function $\delta[(k_x-k(\Omega)\text{sin}\gamma)z/2]$. Doing so yields the following closed-form expressions for conversion efficiency for the DG-TPF case,

\begin{gather}
\eta = Le^{-\frac{\alpha L}{2\text{cos}\gamma}}\bigg(\frac{J_{pump}{\chi^{(2)}}^2n_{THz}}{2\pi n_{g}^2n_{NIR}^2c^3\varepsilon_0}\bigg)\frac{1}{\tau^3}\left(1-\frac{3k_T^2w^2}{\tau^4}\right)~, |k_T|w/\tau^2 \ll 1,\label{tpf_eta_a}\\
\eta = Le^{-\frac{\alpha L}{2\text{cos}\gamma}}\bigg(\frac{J_{pump}{\chi^{(2)}}^2n_{THz}}{2\pi n_{g}^2n_{NIR}^2c^3\varepsilon_0}\bigg)\frac{\tau}{|k_T|w^2} ~, |k_T|w/\tau^2 \gg 1.\label{tpf_eta_b}
\end{gather}

Equation (\ref{tpf_eta_a}) considers the case when GVD-AD is small ($|k_T| w/\tau^2 \ll 1$), similar to an expression derived in \cite{ravi2018}. Notably, the scaling with $k_Tw\tau^{-2}$ in Eq. (\ref{tpf_eta_a}) is identical to the expression in \cite{ravi2018} while the prefactor is different due to a slightly modified derivation. In Eq. (\ref{tpf_eta_a}), $L$ is the propagation length in the $z$ direction, while $J_{pump}$ is the pump energy per unit length, given by

\begin{equation}
J_{pump}=\pi c\varepsilon_0 n_{THz} \int_0^\infty\int_{-\infty}^{\infty}|E(\Delta\omega),x|^2d\Delta\omega dx.
\end{equation}
\noindent In comparing terahertz generation by DG-TPF's and beamlet superposition, we assume $J_{pump}$ to be identical
in this section.

In Eq. (\ref{tpf_eta_a}), $\eta$ shows an optimal duration $\tau_{opt} = 1.62(|k_T|w)^{1/2}$. For $w= 5\,$mm, this corresponds to $\approx 413\,$fs, which is comparable to earlier theoretical predictions. Note that $\tau_{opt}$ increases for larger beam sizes, which translates to lower generated terahertz frequencies. In the total absence of GVD-AD (i.e. $k_T=0$), the efficiency scales inversely as the pulse duration cubed or $\tau^{-3}$. In contrast, in the regime where $|k_T|$ is very large, the efficiency shows the opposite trend and reduces with $\tau$. Furthermore, the efficiency drops with $w^2$ in for high $|k_T|$. These equations justify earlier calculations and prove that terahertz generation using DG-TPF's deteriorates for larger pump beam sizes and bandwidths.

In contrast, the beamlet superposition case is closer to the case of a disortion-free tilted pulse front (i.e for $k_T=0$ in Eq. (\ref{tpf_eta_a})),
\begin{gather}
\eta = Le^{-\frac{\alpha L}{2\text{cos}\gamma}}\bigg(\frac{J_{pump}{\chi^{(2)}}^2n_{THz}}{2\pi n_{g}^2n_{NIR}^2c^3\varepsilon_0}\bigg)\frac{1}{(\tau^2+(\sigma\text{sin}\gamma/v_{THz})^2)^{3/2}}.\label{ech_eta}
\end{gather}

Equation (\ref{ech_eta}) suggests an effective pulse duration is given by $\sqrt{\tau^2 + (\sigma\text{sin}\gamma/v_{THz})^2}$, which increases with larger $\sigma$. This implies that $\sigma$ has to be sufficiently small to allow for sufficient transverse momentum spread to generate the necessary terahertz radiation, identical to the threshold condition deduced in Eq. (\ref{beam_th}). Furthermore, it is evident that the conversion efficiency due to beamlet superposition will always be lesser than a distortion-free tilted pulse front.

If one takes the ratio of Eq. (\ref{ech_eta}) and Eq. (\ref{tpf_eta_a}) at the threshold condition for $\sigma$ given by Eq. (\ref{beam_th}), one obtains the following conditions for efficiency parity, corresponding to the two limiting cases for gratings outlined in Eqs. (\ref{tpf_eta_a})-(\ref{tpf_eta_b}) respectively,

\begin{gather}
\frac{|k_T|w}{\tau^2}\geq 0.46,\, |k_T|w/\tau^2 < 1, \nonumber\\
\frac{|k_T|w}{\tau^2}\geq 1.18, \, |k_T|w/\tau^2 > 1. \label{eff_parity}	
\end{gather}

This shows quite generally that beamlet superposition offers a much better performance for large bandwidths and beam sizes.

\subsection{Optimizing efficiency}
Obtaining global optima requires detailed numerical simulations accounting for cascading effects. However, some general bounds for parameters can be established analytically: 
\begin{enumerate}
    \item The pulse duration $\tau$ must contain sufficient bandwidth to generate the desired frequency according to the relation $f_{THz}=1/(\pi\tau)$;
    \item The optimal crystal length will either be limited by the terahertz absorption length or the length over which material dispersion spreads the pulse to a duration much larger than its transform limit; and
    \item The beamlet size $\sigma$ has to be small enough to provide the necessary angular spread to generate $f_{THz}$ but large enough to circumvent limitations of either pump diffraction or terahertz diffraction.
\end{enumerate}  
\noindent Items (2) and (3) can be expressed as,

\begin{gather}
L = \textbf{min}\bigg( \frac{2\text{cos}\gamma}{\alpha},\frac{\tau^2}{2k_m}\bigg),\label{L_min}\\
\sigma^2 \geq  \textbf{max}\bigg(\frac{\lambda_0 L}{2\pi n(\lambda_0)},\frac{\lambda_{THz}\Delta x}{\pi n_{THz}\text{sin}\gamma}\bigg).\label{sigma_min}
\end{gather}

\noindent The upper bound on condition (3) was already outlined in Eq. (\ref{beam_th}). For pump diffraction, the Rayleigh range has to be larger than the optimal interaction length outlined in Eq. (\ref{L_min}). However, for terahertz diffraction, the terahertz Rayleigh range only needs to be larger than $\Delta x/\text{sin}\gamma$, which is the distance between adjacent pump beamlets.

\begin{figure}
\begin{center}
\scalebox{0.4}[0.4]{\includegraphics{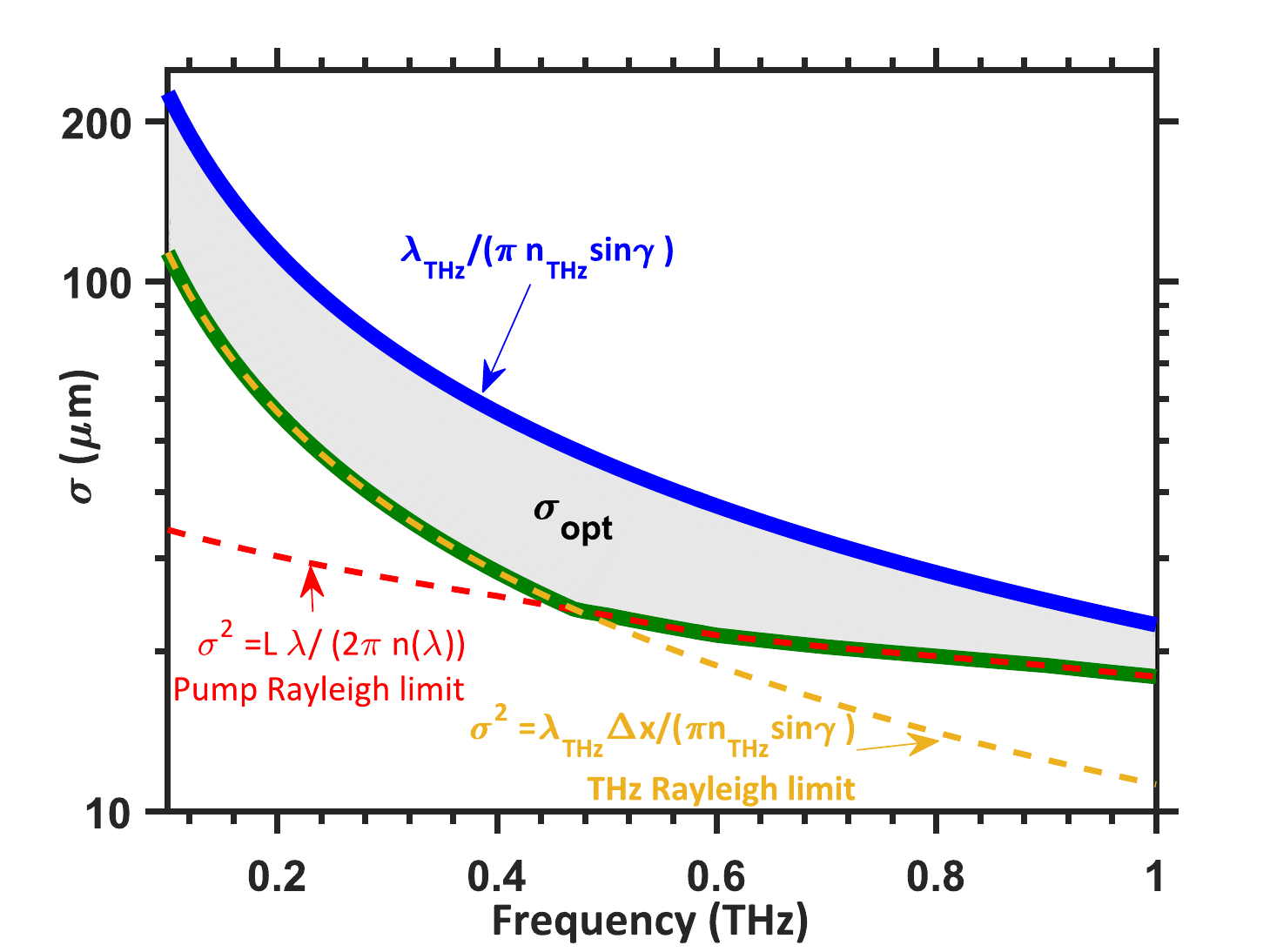}}
\caption{\label{fig14} Optimal parameters for beamlet superposition for lithium niobate. The optimal pulse duration is related to the central terahertz frequency by $f_{THz}=(\pi\tau)^{-1}$. The maximum length is absorption limited and given by $L =2\text{cos}\gamma/\alpha$. The upper limit of beamlet radius $\sigma$ is given by Eq. (\ref{beam_th}), while the lower limits are determined either by terahertz or pump beamlet diffraction as shown in Eq. (\ref{sigma_min})}
\end{center}
\end{figure}

We plot the bounds for beamlet size in Fig. \ref{fig14} for cryogenically cooled lithium niobate as a function of $f_{THz}$. For the case of lithium niobate, the interaction length $L$ is primarily limited by absorption at all values of $f_{THz}$. The maximum value of $\sigma$ is given by Eq. (\ref{beam_th}) while assuming $\Delta x=\sigma/2$, while the minimum value is given by Eq. (\ref{sigma_min}). The optimum lies within the shaded region and as stated previously, will require detailed simulations. For the same reason, cases where the continuous reduction of $\sigma$ produces detrimental effects are not discussed in the current work.  At lower terahertz frequencies, terahertz diffraction is the main limitation. For larger terahertz frequencies, one is limited by diffraction of the pump beam due to the need for smaller beamlet sizes.

\section{Conclusion}

We were able to show by a combination of analytic and semi-analytic methods, the advantages of terahertz radiation generated by beamlet superposition over tilted pulse fronts generated by diffraction gratings for large pump beam radii and bandwidths. Closed-form spatio-temporal expressions enabled us to shed light on the general physics of terahertz generation using this approach. Conditions for obtaining parity in efficiency were furnished and bounds for optimal parameters were provided. 

While we dealt with purely undepleted models, the important finding was that beamlet superposition performs better in relation to DG-TPF's when overall beam sizes or bandwidths are larger. In effect, cascading is just an increase in the bandwidth of the pulse and hence the improvement in performance at larger bandwidths signals alleviation of cascading limitations. While initial simultations appear to attest to this hypothesis \cite{ravi2016}, future work will address the ultimate limits of beamlet superposition.

\section{Appendix}
\begin{table*}[h!]
\captionsetup{labelfont=bf}
\centering
\caption{\label{var_list} \textbf{List of variables} }
\scalebox{0.7}{
\begin{tabular}{l c r}
\hline
Symbol& Variable & Units (SI) \\
\hline
$\omega$ & Optical angular frequency & rads/s \\

$\omega_0$ & Central optical  angular frequency & rads/s \\

$\Delta\omega$ & Displacement from $\omega_0$&rads/s\\

$\gamma$  & Tilt angle & rads \\

$\Delta x$ & Transverse separation between adjacent beamlets & m \\

$\Delta z$ & Longitudinal separation between adjacent beamlets & m \\

$\Delta t$ & Temporal separation between adjacent beamlets & s \\

$k_0$ & Central optical wave number & m $^{-1}$ \\

$k(\omega)$ & Optical wave number  & m$^{-1}$   \\

$w_y$ & Input beam radius in the $y$-direction & m\\

$\phi^{(2)}$ & Input group delay dispersion & s$^2$\\

$v_g$& Optical group velocity & m/s\\

$n_{g}$& Optical group refractive index & \\

$\tau$ & Input pump pulse duration & s\\

$\tau_1$ & Effective pump pulse duration & s\\

$\Omega$ & Terahertz angular frequency & rads/s\\

$k(\Omega)$ & Terahertz wave number  & m$^{-1}$   \\

$\sigma(z)$ & z-dependent beamlet radius& m\\

$\sigma$ & Minimum beamlet radius& m\\

$R(z)$ & z-dependent radius of curvature & m\\

$z_R$ & Rayleigh length & m\\

$z_0$ & Beam waist position & m\\

$k_m$& Group velocity dispersion due to material dispersion&  s$^2$m$^{-1}$\\

$\beta$ & Angular dispersion & rads-s\\

$w$ & Input beam radius in the $x$-direction & m\\

$v_{THz}$& Terahertz phase velocity & m/s\\

$n_{THz}$& Terahertz phase refractive index & \\

$\Delta k$&Phase mismatch& m$^{-1}$\\

$\beta''$ &  Total group delay dispersion & s$^2$\\

$w_1$ & Effective beam radius in the $x$-direction & m\\

$k_T$& Group velocity dispersion due to angular dispersion&  s$^2$m$^{-1}$\\

$x_{p}$& $p^{th}$ beamlet position in the $x$-direction & m \\   

$z_{p}$& $p^{th}$ beamlet position in the $z$-direction & m \\   

$\langle x_{p,q}\rangle$ & $(x_p+x_q)/2$ & m \\   

$\langle z_{p,q}\rangle$ & $(z_p+z_q)/2$ & m \\

$\Delta x_{p,q}$ & $x_p-x_q$ & m \\

$\Delta z_{p,q}$ & $z_p-z_q$ & m \\

$E_{max}$ & Peak electric field for beamlet superposition & Vm$^{-1}$ \\

$\sqrt{2\pi} E_{0}/\tau$ & Peak electric field for DG-TPF's & Vm$^{-1}$\\
\hline
\end{tabular}}
\end{table*}

The following conventions are adopted for Fourier and inverse Fourier operations in the time domain.

\begin{gather}
f(\omega) = \frac{1}{2\pi}\int_{-\infty}^{\infty}f(t)e^{-j\omega t}dt ~,~
f(t) = \int_{-\infty}^{\infty}f(\omega)e^{j\omega t}d\omega 
\end{gather}\label{FT_t}

The following conventions are adopted for Fourier and inverse Fourier operations in the spatial domain.

\begin{gather}
f(k_x) = \frac{1}{2\pi}\int_{-\infty}^{\infty}f(x)e^{jk_x x}dx ~,~
f(x) = \int_{-\infty}^{\infty}f(k_x)e^{-jk_x x}dk_x\label{FT_X}
\end{gather}

The representation of Parseval's theorem or energy conservation is as follows:
\begin{gather}
\int_{-\infty}^{\infty}|f(x))|^2dx = 2\pi\int_{-\infty}^{\infty}|f(k_x)|^2dk_x \label{parseval}
\end{gather}

The nonlinear polartization, $P_{THz}(\Omega,x,y,z)$, that drives terahertz generation is calculated using the pump field $E(\omega,x,y,z)$ as follows:
\begin{gather}
P_{THz}(\Omega,x,y,z) =\int_0^\infty E(\Delta\omega+\Omega,x,y,z)E^*(\Delta\omega,x,y,z)d\Delta\omega    
\end{gather}

The general three-dimensional scalar wave equation after Fourier decomposition can be presented as follows:
\begin{gather}
\nabla^2 E_{THz}(\Omega,k_x,k_y,z)+k_z^2(\Omega)E_{THz}(\Omega,k_x,k_y,z) = -\frac{\Omega^2}{\varepsilon_0c^2}P_{THz}(\Omega,k_x,k_y,z) \label{wav_eq_k}
\end{gather}
The solution to the above equation maybe expressed by the ansatz $E_{THz}(\Omega,k_x,k_y,z)=A_{THz}(\Omega,k_x,k_y,z)e^{-jk_z.z}$. Using the above ansatz and integrating over $z$ , one obtains the following expression.
\begin{gather}
A_{THz}(\Omega,k_x,k_y,z) = -\frac{j\Omega^2\chi^{(2)}\int_0^z P_{THz}(\Omega,k_x,k_y,z)e^{j\Delta kz+\frac{\alpha z}{2\text{cos}\gamma}}dz}{2k_z(\Omega)c^2}\label{eq_a}
\end{gather}

One may deal with the many $z$ dependent terms in the polarization term as follows.  If one has an integral of the form $\int P(z)e^{j\Delta kz +\frac{\alpha}{2}z}dz$, it may be reduced to Eq. (\ref{int_by_parts}) upon using integration by parts (i.e. $\int u(z)v'(z)dz = uv -\int u'(z)v(z)dz$).
\begin{gather}
\int P(z)e^{j\Delta kz +\frac{\alpha}{2}z}dz\nonumber\\
=\frac{P(z)e^{j\Delta kz +\frac{\alpha}{2}z}}{\alpha/2+j\Delta k}-\frac{P'(z)e^{j\Delta kz +\frac{\alpha}{2}z}}{(\alpha/2+j\Delta k)^2}+\frac{P"(z)e^{j\Delta kz +\frac{\alpha}{2}z}}{(\alpha/2+j\Delta k)^3}\label{int_by_parts}
\end{gather}

For typical absorption coefficients and $\Delta k$ values, the second terms and beyond are much smaller and can be ignored. Therefore, we have $\int P(z)e^{j\Delta kz +\frac{\alpha}{2}z}dz = P(z)e^{j\Delta kz +\frac{\alpha}{2}z}(\alpha/2+j\Delta k)^{-1}$.

Using the above result,  expanding $k_z$ about $k(\Omega)\text{cos}\gamma$ (since this is the direction in which phase-matched terahertz radiation propagates) and performing an inverse Fourier operation on Eq. (\ref{eq_a}) in the spatial domain, one obtains Eq. (\ref{e_r_z_g}) under the condition that $\alpha\gg\Delta k$. 

\begin{gather}
E_{THz}(\Omega,x,y,z) = \frac{-j\Omega^2\chi^{(2)}}{\alpha k(\Omega)c^2}\bigg[P_{THz}(x,y,z) - \frac{e^{-\frac{\alpha z}{2\text{cos}\gamma}}}{4\pi^2}\nonumber\\
\times \int\int P_{THz}(\Omega,k_x,k_y,z)e^{-jk_yy}e^{j\frac{k_y^2z}{2k(\Omega)\text{cos}\gamma}}e^{-jk_xx}e^{j\frac{k_x^2z}{2k(\Omega)\text{cos}\gamma}+k_xz\text{tan}\gamma}dk_xdk_y\bigg]\label{e_r_z_g}
\end{gather}

The first term denotes the nonlinear polarization term and the second term represents the field that it radiates. The radiation term is nothing but a Fresnel integral with modified kernel to account for oblique propagation. Therefore, the first term in Eq. (\ref{e_r_z_g}) is referred to as the source (or near-field) term while the second term is referred to as the propagation (or far-field) term.

The conversion efficiency may be calculated as follows :

\begin{gather}
\eta(z) = \frac{1}{\sqrt{2}}2\pi^2 c\varepsilon_0n_{THz}\frac{\int_0^{\infty}\int_{-\infty}^{\infty}|E(\Omega,k_x,z)|^2dk_xd\Omega}{F_{pump}w\sqrt{\pi/2}}
\end{gather}

The $1/\sqrt{2}$ pre-factor arises from the averaging effect in the $y$-dimension.

\section*{Funding}

 Air Force Office of Scientific Research (AFOSR) (A9550-12-1-0499); European Research Council (ERC, FP/2007-2013) (n.609920); Center for Free-Electron Laser Science (CFEL).
 
\section*{Acknowledgment}

The authors thank Erich P. Ippen and Phillip Donald Keathley for their assistance in improving the manuscript and Prasahnth Sivarajah and Wenqian Ronny Huang for discussions.

\bibliography{sample}
\bibliographystyle{ieeetr}





\end{document}